\documentclass[twocolumn,english]{revtex4}
\usepackage{color}
\usepackage{bm}
\usepackage{amsthm}
\usepackage{amsmath}
\usepackage{esint}
\usepackage{graphicx,epsf}
\usepackage{dcolumn}
\usepackage{amsfonts}
\usepackage{mathrsfs}
\usepackage{subfig}
\usepackage{mathrsfs}
\usepackage{bbding}

\makeatletter
\begin{document}

\title{The enhanced soliton propagation and energy transfer in the coupled drift wave and energetic-particle-induced geodesic acoustic mode system}

\author{Ningfei Chen$^1$, Guangyu Wei$^1$, and Zhiyong Qiu$^{1,2}$}

\affiliation{$^1$Institute for  Fusion Theory and Simulation, School of Physics, Zhejiang University, Hangzhou, P.R.C\\
$^2$ Center for Nonlinear Plasma Science and   C.R. ENEA Frascati, C.P. 65, 00044 Frascati, Italy}

\begin{abstract}
The evolution of the coupled drift wave (DW) and energetic-particle-induced
geodesic acoustic mode (EGAM) nonlinear system is investigated   using
the fully nonlinear coupled  DW-EGAM two-field equations, with emphasis on the turbulence spreading in the form of soliton and the nonlinear energy transfer between DW and EGAM. Four scenarios with different combinations of EGAM initial amplitudes and linear EGAM growth rates are designed to delineate the effects of linear EGAM drive and finite EGAM amplitude on DW nonlinear dynamic evolution.
In presence of the linear EPs drive, the soliton propagation is enhanced,  due to the generation of small radial scale structures. Two conservation laws of the nonlinear
system are derived, including the energy conservation law.   It is found that the energy of
 DW always decreases and that of EGAM always increases,
leading to regulation of DW by EGAM.

\end{abstract}
\maketitle

\section{Introduction}

The anomalous transport, generally accepted to be triggered by
micro-turbulence \cite{WHortonRMP1999}, is a major concern for tokamak confinement \cite{KTomabechiNF1991}.
The micro drift wave (DW) turbulences can
be driven unstable by plasma pressure gradient intrinsic to magnetically confined plasmas, and are, thus,
ubiquitous modes in magnetically confinement devices \cite{KTomabechiNF1991,YWanNF2017}. To achieve
better performance, the DW turbulence should be regulated to lower transport level. It is not only the amplitude of the DW in its
unstable region that matters, but also the radial redistribution of
wave intensity due to radial propagation of the DW envelope
from its linearly unstable to stable region, i.e., turbulence spreading
\cite{XGarbetNF1994}. Previous investigations suggest that turbulence
spreading contributes to the nonlocal transport and core-edge coupling,
i.e., turbulent transport is dependent on the plasma parameters
elsewhere \cite{TSHahmPPCF2004,HahmPoP2005}. Zonal flows (ZFs) \cite{AHasegawaPoF1979,PDiamondPPCF2005} are found to play important roles in both regulating DW turbulence via shearing \cite{ZLinScience1998} and inducing turbulence spreading \cite{ZGuoPRL2009,NChenPPCF2022}, and are thus, of crucial importance for turbulence dynamics.

ZFs are radial electric field with $n=0$ and $m\simeq0$ electric
potential, which consist of  zero-frequency ZF (ZFZF) \cite{MRosenbluthPRL1998,PDiamondPPCF2005} and finite
frequency geodesic acoustic mode (GAM) \cite{NWinsorPoF1968,FZoncaEPL2008,ZQiuPST2018,GConwayNF2021}. Here, $n$ and $m$ are, respectively, the toroidal and poloidal mode numbers.
On the one hand,  ZFs are expected  and shown to reduce the turbulence level
by the spatial decorrelation due to  flow shearing, turning the
large eddies into smaller ones and reducing DW amplitude \cite{ZLinScience1998,LChenPoP2000}. The  DW regulation is achieved by spontaneous excitation of ZFs, which in turn, scatters DWs into linearly stable short  wavelength domain.   The excitation of finite frequency counterpart of
ZF, i.e., GAM,  by  DW was also observed in experiments \cite{KZhaoPRL2006,GConwayPPCF2008,TLanPoP2008,WZhongNF2015,AMelnikovNF2017},  which also leads to confinement improvement.
Due to the finite frequency
of GAM, it can be excited by velocity space anisotropy of
energetic particles (EPs), i.e., the EPs-induced GAM (EGAM), as reported  and analyzed in Refs.  \citenum{RNazikianPRL2008,GFuPRL2008,ZQiuPPCF2010}. EGAM is typically characterized with a frequency lower than local GAM frequency, and meso/macro mode structure determined by radial profile of EP intensity.
Thus, EGAM excitation by externally injected EPs is proposed as a potential active control of  turbulence \cite{ZQiuPST2018}.  However, it was observed in a recent gyrokinetic simulation  that,    DW can be excited from the Dimits-shifted  marginally stable region
by  EGAM, indicating possible energy flow from
 EGAM to  DW \cite{DZarzosoPRL2013}.  To understand this unexpected ``excitation'' of DW turbulence by EGAM, a recent work was carried out by studying the    linear growth rate of the  ion-temperature gradient (ITG) turbulence under the influence of the equilibrium radial electric field  with the frequency being much lower than that of ITG \cite{NChenPoP2021}, i.e., the nonlinear equilibrium. It was shown, however,  that the ITG linear growth rate is reduced by
equilibrium radial electric field in both short- and long-wavelength limit, align with
the common speculation of the DW turbulence suppression by  ZFs.  But nonlinear interactions between   DW and   EGAM has not been
investigated, especially nonlinear energy transfer of the nonlinear
coupled system.

On the other  hand, ZFs are suggested to be the mediator
for the formation of DW-ZFs solitons, resulting in turbulence
spreading \cite{ZGuoPRL2009,NChenPPCF2022}.
The fully nonlinear  DW-ZFs
interaction was firstly investigated by Ref. \cite{ZGuoPRL2009}, in which the balance of
 DW dispersiveness and nonlinear trapping effect due to   ZFZF
results in the formation of the DW-ZFZF soliton. An important breakthrough
in that work was the derivation of the coupled DW-ZFZF two-field model, which
does not separate the DW into fixed-amplitude pump wave and relatively small amplitude sidebands, as usually  done in the parametric decay/modulational instability analysis \cite{LChenPoP2000,FZoncaPoP2004}.
The two-field model is thus, able to describe the long time evolution of
the  coupled DW-ZFs   system. This novel approach was then extended to
the coupled DW-GAM system, and the formation of the DW-GAM solitons was observed in
the nonlinear saturated stage.
Thus,
the investigations on the DW-EGAM nonlinear system is a natural and
necessary extension of the topic, i.e., nonlinear mechanisms for DW turbulence spreading, besides its apparent relevance to understand the results of Ref. \citenum{DZarzosoPRL2013}.

Motivated by the above-mentioned advances, in this work, the nonlinear
interaction between   DW and EGAM is systematically investigated
using the coupled DW-EGAM two-field model, based on the nonlinear gyrokinetic
framework. Four scenarios are considered in this work to delineate the effects of linear EP drive and nonlinear mode coupling, which are the
combination of with/without linear EPs drive and the initially small/finite
amplitude EGAM.  Here, for ``small", we mean the GAM/EGAM were excited from noise level, as will be shown later in Sec. \ref{sec:Theoretical-model}.    The formation and propagation of the DW-EGAM solitons
are investigated for the four scenarios, and the enhanced DW-EGAM soliton velocity
is found,   compared to that without the linear
EPs drive \cite{NChenPPCF2022}, indicating that the linear EPs drive
might enhance the turbulence spreading. The spectrum of the DW and
GAM/EGAM are investigated, and it is found that the accelerated propagation
might be related to the excitation of micro-scale DW (high-$k_{r}$
harmonics, with $k_r$ being the radial wavenumber).   In terms of the energy transfer, two conservation laws
of the coupled nonlinear DW-EGAM system are derived, including the energy
conservation law \cite{VZarharovJETP1972}. Both of them are verified numerically, which are shown to be
conserved to accuracy required by the investigations on the energy
transfer. Total energy can be decomposed into the energy of DW, GAM/EGAM, and interaction.
It is found that the energy of DW always decreases while that of
 EGAM/GAM always increases in all  four scenarios, so the unexpected ``excitation'' of DW turbulence by EGAM \cite{DZarzosoPRL2013} does not happen in the present system. Thus,
this work offers new insight to the nonlinear DW suppression
through the energy transfer to EGAM.

The rest of this paper is organized as follows: in  Sec. \ref{sec:Theoretical-model},   the coupled DW-EGAM two-field equations are derived based on the nonlinear
gyrokinetic theoretical framework. The soliton propagation and associated spectrum evolution
of DW and GAM/EGAM are investigated in  Sec. \ref{sec:Accelerated-propagation}. Two conservation
laws are derived in Sec. \ref{sec:Energy-flow}, and the evolution of each energy components
is also looked into to understand the energy transfer between DW and EGAM. Finally, the conclusion and discussion are presented in
Sec. \ref{conclusion}.

\section{Theoretical model\label{sec:Theoretical-model}}

In this paper, we study the nonlinear interaction between   DW and
EGAM in a large aspect ratio tokamak using the nonlinear gyrokinetic
theory \cite{EFriemanPoF1982}. The detailed  derivation of the coupled nonlinear DW-EGAM equations follow closely  that in   Refs.
\cite{FZoncaEPL2008,NChenPPCF2022}, with additional twists  focusing
on the proper  inclusion  of   EPs. For simplicity of discussion while focusing on the main scope,
 DW and EGAM are both taken as  electrostatic fluctuations, satisfying proximity
to marginal stability, i.e. $\left|\gamma_{L}/\omega\right|\ll1$,
to assure that the parallel mode structure of DW is not disturbed in
the relatively weak nonlinear envelope modulation process. Here, $\gamma_L$ and $\omega$ are, respectively, mode linear growth rate and real frequency.  Neglecting the weak influence of EPs on the DWs with predominantly $k_{\perp}\rho_i\sim O(1)$ \cite{WHortonRMP1999}, the nonlinear evolution equation
of  DW remains unchanged.  Here, $k_{\perp}$ is the perpendicular wavenumber, and $\rho_i$ is the ion Larmor radius.    On the other hand, the nonlinear  equation for the
GAM/EGAM should be properly derived, noting that, EPs enter through linear resonant interaction with GAM \cite{GFuPRL2008,ZQiuPPCF2010}, while nonlinearity is dominated by thermal ion contribution to Reynolds stress \cite{FZoncaEPL2008}. Separating the  nonadiabatic particle responses
  into linear and nonlinear components $\delta H=\delta H^{L}+\delta H^{NL}$,  with the superscripts ``L" and ``NL" denoting linear and nonlinear responses, respectiely, the quasi-neutrality
condition for EGAM can be written as
\begin{eqnarray}
 & \dfrac{e^{2}N_{0}}{T_{i}}(1+\dfrac{1}{\tau})\delta\phi-\left\langle eJ_{0}\delta H_{i}^{L}\right\rangle +\left\langle e\delta H_{e}^{L}\right\rangle -\left\langle eJ_{0}\delta H_{EP}^{L}\right\rangle \nonumber \\
 & =\sum_{s=i,e}\left\langle eJ_{0}\delta H_{s}^{NL}\right\rangle. \label{eq:QN}
\end{eqnarray}
Here, $J_k\equiv J_0(k_\perp\rho_{s})$ is the Bessel function of zero index describing the finite Larmor radius effect (FLR)  with $\rho_{s}$ being the Larmor radius for species $s$,  $\tau\equiv T_e/T_i$ is the temperature ratio,   $N_0$ is the number of particles, and the nonadiabatic particle response, $\delta H$, is derived from the nonlinear gyrokinetic equation \cite{EFriemanPoF1982}
\begin{eqnarray}
 & \left(\omega-k_{\parallel}v_{\parallel}+\omega_{D,s}\right)\delta H_{k,s}=\dfrac{e_s}{T_{s}}J_{0}\left(k_{\perp}\rho_{s}\right)\left(\omega+\omega_{*,s}\right)F_{0,s}\nonumber \\
 & -i\Lambda_{\mathbf{k'},\mathbf{k''}}^{\mathbf{k}}J_{k'}\delta H_{k''}\delta\phi_{k'}.\label{eq:NLGKE}
\end{eqnarray}
Here, $\omega_{*,s}= k_\theta cT_s(1+\eta_s(v^2/v_{ts}^2-3/2))/(e_sBL_{n})$ is the diamagnetic
frequency accounting for both gradients of equilibrium density and temperature profiles,
 $\omega_{D,s}= \mathbf{k}\cdot(\mathbf{B}\times\nabla \mathbf{B}/B^2)(v_\perp^2/2+v_\parallel^2)/\omega_{c,s}$ is the magnetic drift frequency, with $\omega_{c,s}=e_sB/(m_sc)$ being the cyclotron frequency for species $s$, and $\eta_s\equiv L_{n}/L_{T,s}$ is the ratio of the characteristic length of density variation $L^{-1}_{n}\equiv -\partial\ln(n)/\partial r$ and temperature variation $L^{-1}_{T,s}\equiv-\partial\ln(T_s)/\partial r$. The last
term is the formal nonlinear term, with $\Lambda_{\mathbf{k'},\mathbf{k''}}^{\mathbf{k}}\equiv(c/B_0)\sum_{\mathbf{k}=\mathbf{k'}+\mathbf{k''}}\mathbf{b}\cdot{\left(\mathbf{k''}\times\mathbf{k'}\right)}$ indicating selection
rule for mode-mode coupling and the subscript `$k$' represents the quantities associated with mode $\mathbf{\Omega}_k(\omega_k, \mathbf{k})$.
Substituting the obtained particle responses into quasi-neutrality condition, one obtains, the governing two field  equations describing DW and EGAM/GAM nonlinear interactions:
\begin{eqnarray}
&&
\omega\mathscr{E}_du=\dfrac{ck_\theta}{\tau B_0}uv,\label{eq:DW_1}\\
& & \omega\mathscr{E}_Ev=-\alpha_i\dfrac{ck_\theta}{B_0}(u\partial_r^2u^*-c.c.).\label{eq:GAM_1}
\end{eqnarray}

Here, $u$ and $v$ are the radial envelope of the DW electrostatic potential and the electric field of the EGAM/GAM, respectively,   $\alpha_{i}$ is an order  unity factor related to the perturbed ion pressure  \cite{LChenPoP2000}, and $k_\theta\equiv m/r$ is the poloidal wavenumber. Furthermore,  $\mathscr{E}_{d}$ and $\mathscr{E}_{E}$  are the linear dispersion relation of  DW and EGAM/GAM, respectively.    The DW dispersion relation   adopted in this work is, again, that for the electron DW $\mathscr{E}_{d}=1-\omega_*/\omega+C_d(\omega_*/\omega)k_r^2\rho_i^2$, with $\omega_{*}\equiv k_{\theta}cT_{i}/(eBL_{n})$ being the ion diamagnetic frequency,  $C_d$ representing the kinetic dispersiveness of DW \cite{FRomanelliPoFB1993}.
  The nonlinear terms on the right-hand-side (RHS) of equation (\ref{eq:DW_1}) represent
 the modulation of the DW envelope by GAM/EGAM.

 More attention should be paid to equation (\ref{eq:GAM_1}), with the left-hand-side (LHS) describing GAM/EGAM linear properties, while the RHS being nonlinear drive by DW Reynolds stress \cite{FZoncaEPL2008,ZQiuPoP2014}.  For the typical slowing down circulating EPs distribution function, the linear  EGAM dispersion relation  can be written as $\mathscr{E}_{E}=-1+\omega_{E}^{2}/\omega^{2}+C_{G}(\omega_{E}^{2}/\omega^2)k_{r}^{2}\rho_{i}^{2}+C\ln(1-\omega_{tr,b}^{2}/\omega^{2})+D/(\omega^{2}/\omega_{tr,b}^{2}-1)$
\cite{ZQiuPPCF2010}. Here, $C_G$, $C$ and $D$ are coefficients defined in  Ref. \cite{ZQiuPPCF2010}, $\omega_{tr,b}$ is the transit frequency of the EPs at birth energy. The 4th and 5th terms correspond to the resonant
EPs drive and EP induced real frequency shift, with the former contributing to  the crucial EGAM resonant excitation  \cite{GFuPRL2008,ZQiuPPCF2010}.  The GAM/EGAM dispersion relation, $\mathscr{E}_E$, yields two main branches,  i.e., GAM branch and beam branch \cite{ZQiuPPCF2010}, with the  former   being essentially local GAM with perturbative contribution from EPs, while the latter is an EP mode  with the frequency and mode structure dominated by resonant EP drive,  analogous to the plasma wave and beam mode in the well-known beam-plasma instability  \cite{TOneilPoF1971}.  Note that, though the GAM   and beam branches have distinctive properties, they can both be well described by the above dispersion relation.
For the GAM branch, the $\omega_{E}$ shown above can be straightforwardly  understood
as the local GAM frequency  with the lowest order expression being $\omega_{G}^{2}\equiv(7/4+\tau)v_{ti}^{2}/R^{2}$;
 while for the non-perturbative beam branch with the frequency dominated by EP transit frequency, one has  $\omega_{E}\approx\omega_{tr,b}$. Since both GAM and EGAM interaction with DWs are investigated in this work, $\omega_G$ will be used in the rest of manuscript to simplify the notation.

Employing the definitions above, and expanding $\mathscr{E}_E$ and $\mathscr{E}_d$ along the characteristics, the DW-EGAM two-field
equations can be cast into
\begin{eqnarray}
 & u_{t}+iu+iC_{d}u_{rr}=-i\Gamma_{0}uv,& \label{eq:DW Eq}\\
& v_{tt}-2\gamma_{G}v_{t}+\omega_{G}^{2}(v-C_{G}v_{rr})=i\alpha_{i}\tau\Gamma_{0}(uu_{rr}^{*}-c.c.)_{t}, & \label{eq:GAM Eq}
\end{eqnarray}
which correspond to   equations (6) and (7) of Ref. \cite{NChenPPCF2022}, with the additional physics of EP drive to GAM via finite $\gamma_G$. Here, $\Gamma_{0}$ is the nonlinear coupling
coefficient,  $C_{d}$ and $C_{G}$, again,  represent the kinetic dispersiveness
of DW and EGAM \cite{FRomanelliPoFB1993,ZQiuPPCF2009},    and other notations used here are standard.
 The
set of equations can describe the fully nonlinear DW-EGAM dynamics, i.e.,
the self-modulation of DW via GAM excitation,  and the generation of   soliton structures that influence the turbulence spreading \cite{NChenPPCF2022}.   For the convenience of
numerical investigation and brevity of notation, space and time are
normalized to $\rho_{i}$ and $\omega_{*}^{-1}$, respectively, i.e.,
$r\rightarrow r/\rho_{i}$, $t\rightarrow\omega_{*}t$, $k_{r}\rightarrow k_{r}\rho_{i}$,
$\omega\rightarrow\omega/\omega_{*}$.  Finally, $u$ and $v$ are normalized to $e/T_{e}$.

Note here that, two mechanisms for EGAM/GAM
excitation exist in this system,  with one being the nonlinear
excitation by Reynolds stress of DW, which mainly pumps energy to the micro-scale
($k_{r}\simeq0.32$ in the present case, determined by the frequency/wavenumber matching
condition for the parameters given in the last paragraph of the present section) and the other  being  the injection of energy to the macro-scale
by the linear EPs drive given at the end of the present section.

In order to thoroughly investigate the effects of the linear
EPs drive on nonlinear EGAM-DW interactions, four scenarios are investigated in this work, corresponding
to combination of with/without linear EPs drive and the initially small/finite
amplitude EGAM. More specifically, they correspond to the interactions between  1. finite amplitude DW
and small amplitude GAM; 2. finite amplitude DW and finite amplitude EGAM with vanishing $\gamma_G$; 3. finite amplitude DW
and small amplitude EGAM; and 4. finite amplitude DW and finite amplitude EGAM with finite $\gamma_G$. Note that,
while GAM and EGAM are both investigated in this paper,   GAM is only used in the first scenario to represent the noise level mode with vanishing $\gamma_G$, corresponding to the case that GAM being spontaneously excited by DW \cite{ZQiuPoP2014,NChenPPCF2022}. Readers interested in the   dynamics of DW-GAM interaction may refer to Ref. \citenum{NChenPPCF2022} for more details.
If the EGAM is strongly driven by the EPs and saturates (due to wave-particle phase space nonlinearity \cite{ZQiuPST2018,ABiancalaniJPP2017}),   the stationary
saturated EGAM interacts with DW nonlinearly, as demonstrated
by the third phase of the Ref. \cite{DZarzosoPRL2013}. This case corresponds
to the second scenario. The third scenario represents small amplitude EGAM being driven simultaneously,   nonlinearly by  DW  as well as linearly by EPs. Finally, the
forth scenario is related to the interaction between the finite amplitude  DW and the
finite amplitude EGAM that still grows.  The initial conditions and the linear growth rate for the four scenarios are listed in the Table. \ref{table1}.

\begin{center}
\begin{table}
\begin{tabular}{|c|c|c|c|}
\hline
 scenario & $u(t=0)$ & $v(t=0)$ & $\gamma_{G}$\tabularnewline
\hline
\hline
first &  $0.006\exp(-r^{2}/30^{2})$ & small ($10^{-6}rand$) & $\times$\tabularnewline
\hline
second & $0.006\exp(-r^{2}/30^{2})$ & finite  ($0.1\exp(-r^{2}/80^{2})$) & $\times$\tabularnewline
\hline
third & $0.006\exp(-r^{2}/30^{2})$ & small ($10^{-6}rand$) & $\checkmark$\tabularnewline
\hline
forth & $0.006\exp(-r^{2}/30^{2})$ & finite ($0.1\exp(-r^{2}/80^{2})$) & $\checkmark$ \tabularnewline
\hline

\end{tabular}
\caption{The detailed expression for initial conditions for $u$ and $v$ are listed. The scenarios that with the linear growth rate $\gamma_G$ are marked with check mark, while, that without are marked with cross.\label{table1}}
\end{table}

\end{center}

The numerical scheme used here is   the \textit{\textcolor{black}{pesudospectral}} method, which
enables us to obtain an accurate result and have a direct access to
the $k_{r}$ spectrum of the waves. Note that, the nonlinear terms
are multiplied in physical space and transformed back to Fourier space
in this work to save computing power.  The employment
of the spectral method in this work is quite natural and convenient,
because the linear growth rate of EGAM is implemented in the Fourier
space ($k_{r}$-space), in order to pump energy exactly to the macro-scale
structures, in consistency with the global nature of EGAM \cite{RNazikianPRL2008}. The practical growth rate is introduced to the system
in the form $\gamma(k_{r})F\{v_{t}\}$, with $F\{\cdots\}$ representing
the Fourier transform of a variable. The outgoing boundary is adopted
to  avoid unphysical
reflection at the boundary. The default parameters used here are $C_{d}=C_{G}=\tau=1$,
the nonlinear coupling coefficient $\Gamma_{0}=50$ to assure that
the nonlinear growth rate of  GAM/EGAM do not exceed its real frequency.
The frequency of GAM/EGAM is much lower than that of  DW, so
that $\omega_{G}/\omega_{*}=0.1$ is taken. Note that the
linear EPs drive typically pumps energy to the macro-scale with $k_{\perp}\rho_h\sim O(1)$,
the linear growth rate of the EGAM can be given as $\gamma_{G}(k)=\gamma_{G0}\exp(-k_{r}^{2}/0.1^2)$,
with   $\gamma_{G0}=0.03$, to make sure little energy
is given to the micro-scale ($k_{r}\simeq0.32$, determined by the frequency/wavenumber
matching conditions for the above parameters \cite{ZQiuPoP2014}). The four scenarios to be
investigated are characterized by different initial conditions for DW and GAM/EGAM, as listed in the Table. \ref{table1}. More specifically, the DW is
defined as $u(t=0)=u_{0}\exp(-r^{2}/L_{P}^{2})$, with $u_{0}=0.006$
and $L_{P}=30$; while, the small and finite amplitude EGAM are defined as $v(t=0)=10^{-6}*rand$
and $v(t=0)=0.1\exp(-r^{2}/80^{2})$, respectively. Here, $rand$
is the random distribution, representing the noise nature of the small amplitude GAM/EGAM. The typical structures of these initial conditions are demonstrated in the Fig. \ref{fig1}.

\begin{center}
\begin{figure}
\includegraphics[scale=0.4]{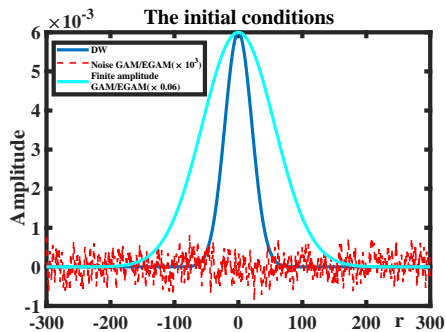}

\caption{Initial conditions of DW and GAM/EGAM. Others are normalized
to the DW initial envelope (blue curve) to have a better view. Here, the blue-green solid line represents the broad mode structure of the EGAM,
the blue solid line is the initial envelope of DW, and the red dashed line shows the random distribution of the small amplitude EGAM/GAM. \label{fig1}}
\end{figure}

\par\end{center}

\section{Accelerated propagation\label{sec:Accelerated-propagation}}

Due to the balance of DW kinetic dispersiveness and nonlinear
trapping effect of  GAM,  it is expected and shown numerically  in Ref. \citenum{NChenPPCF2022} that, DW and GAM  can couple
together and propagate in the radial direction consistently with little dispersion in the form of solitons, causing
 enhanced DW turbulence spreading. Thus, it is naturally interested to find out the effects of
the linear EPs drive on the soliton propagation and also the possible
existence of cross-scale interaction. In align with the previous work
on the spontaneous excitation of GAM by the DW \cite{NChenPPCF2022},
we focus on the comparisons between the first and third scenarios
above, while the second and forth scenarios will also be discussed
briefly.

The initially Gaussian envelope for the DW splits into multiple soliton
structures due to the nonlinear coupling to the EGAM, as shown
in the Fig. \ref{fig2a},  and the velocity
of the solitons are measured by finding the outmost maximums. It is found numerically that propagation
of the coupled DW-EGAM solitons is enhanced. As shown in Fig. \ref{fig3},
the position of the maximum of the DW for linear growth rate of EGAM $\gamma_{G}=0$ and $0.03$ are presented,  and the velocity for
the first and third scenarios are 1.25 and 1.40, respectively. This has an  implication for  fusion devices, that the existence of linear EPs drive may enhance the
turbulence spreading. The lowest order of DW solitons velocity
is $V_{d}=2C_{d}k_{r}$ \cite{ZQiuPoP2014,NChenPPCF2022}, thus, the observed
acceleration might be related to the excitation of higher-$k_{r}$
harmonics. To verify this hypothesis, the $k_{r}$ spectrum and
corresponding spatial-temporal evolution of the DW and EGAM are investigated,
as demonstrated in Fig. \ref{fig4} and Fig. \ref{fig5}.
It is found   that, the $k_{r}$
spectrum of the DW for the third scenario is indeed broader than that for the first scenario, indicating possible relation between the higher-$k_r$ DW spectrum and the accelerated propagation.
The dependence of the velocity on the dominant $k_{r}$ for the first
and third scenarios is not obvious due to the existence of multi-dominant
modes, while the  dependence is more clear for the second scenario shown in the Fig. \ref{fig4c}, in which the mode with $k_{r}\approx1.5$ dominates after $t=100$, corresponding to the single
soliton propagate at $V_{d}\approx3$, as demonstrated in the Fig. \ref{fig2b}.

The reason for the excitation of high-$k_{r}$ modes can be analyzed in a local analysis of the frequency/wavenumber matching condition. The matching condition arises from the parametric decay instability \cite{FZoncaEPL2008,ZQiuPoP2014,NChakrabartiPoP2008}, which correspond to the conservation of the energy and momentum, and can be determined by  $\mathscr{E}_{dr}(\omega_r,k_r)=0$, $\mathscr{E}_{Gr}(\omega_r,k_r)=0$ for maximized drive, especially as the two sidebands are linearly damped. In the present case with finite linear $\gamma_G$, however, the parametric process is thresholdless, thus, a spectrum of modes centered around resonant $k_r\sim0.32$ can excited, and higher-$k_r$ modes with higher growth rates (noting $\gamma\propto k_r$ for resonant decay \cite{ZQiuPoP2014}) and higher propagation velocity may dominate, as observed in the third scenario. Additionally, the enhancement of the resonant $k_r$ is more significant in the second scenario, which is similar to the third scenario, since the growing small amplitude EGAM due to
the linear EPs drive lead to  the macro-scale large amplitude
EGAM. In fact, the second scenario is nearly equivalent to the third scenario with large and transient $\gamma_G$, though the overall growth rate for the EGAM shall not exceed its real frequency in this model. The nonlinear term in the equation (\ref{eq:DW Eq}) for the second scenario is so large ($v(r=0,t=0)=0.1$) that the frequency of DW is strongly increased and the matching conditions are given as $\omega+C_dk_r^2=\Gamma_0v(r=0,t=0)$ and $-\omega^2+\omega_G^2+C_G\omega_G^2k_r^2=0$ \cite{ZQiuPoP2015,NChenPoP2021}.
The lowest order resonant $k_r$ can be derived as $k_r=\sqrt{\Gamma_0v/C_d}$, i.e., $k_r=\sqrt{5}$ for the parameters given in the second scenario, which coincides with the numerical result given in the Fig. \ref{fig4c} ($k_r\sim1.5$). Here, the local assumption is also utilized by treating the EGAM as a stationary background given as $v=v(r=0,t=0)=0.1$, thus, the analysis depicts the resonant condition qualitatively; while radial propagation of the envelope is not covered. The second and third scenarios demonstrate that due to the large amplitude EGAM, the matching condition can be altered, resulting in the increased resonant $k_r$. As a consequence, the rapid expansion of the DW spectrum and
propagation of the DW envelope in the presence of the $\gamma_G$ is reasonable.

Besides, since EGAM is excited in micro- and macro-scales simultaneously by DW and EPs, respectively, the cross-scale interaction intermediated by EGAM is also an important topic. It is shown in the Fig. \ref{fig5a} and Fig. \ref{fig5c}
that EPs pump energy to the macro-scale and DW to the micro-scale
of EGAM. The Fig. \ref{fig5b} and Fig. \ref{fig5c}
are the spectrum of the EGAM/GAM normalized by the maximum at the
corresponding time, in order to find out the dominant radial wavenumber
each time.   When the nonlinearity is turned off, the EGAM with $k_{r}\simeq0$
is excited; while, when the nonlinearity is turned off and in the
absence of the linear EPs drive, the EGAM with $k_{r}\simeq0.32$
is excited, as demonstrated in the Fig. \ref{fig5c}.
If both the nonlinear and linear drives exist, as shown in the Fig. \ref{fig5b}, the spectrum of the EGAM seems to be superposition of the former two cases. More precisely, the nonlinearity dominates
during $t=100-250$, then the EGAM nonlinearly saturates and the linear
EPs drive dominates. Indeed, the interaction
between the macro (linear EPs drive) and the micro-scale (nonlinear
drive) is relatively weak in this nonlinear system, possibly due to the linear nature of the EPs effect in this work.

\begin{center}
\begin{figure}
\subfloat{\includegraphics[scale=0.25]{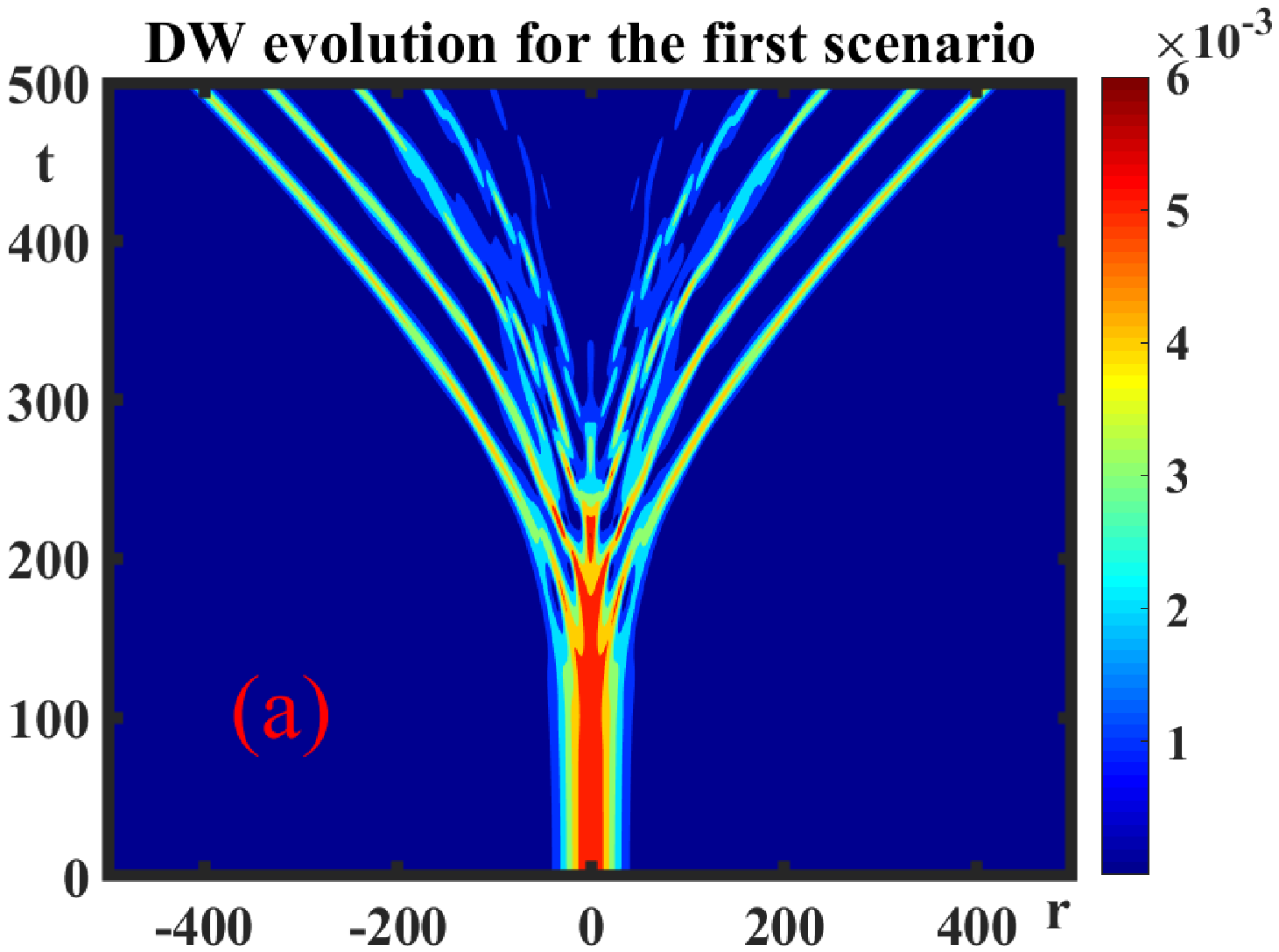}\label{fig2a}}\subfloat{\includegraphics[scale=0.25]{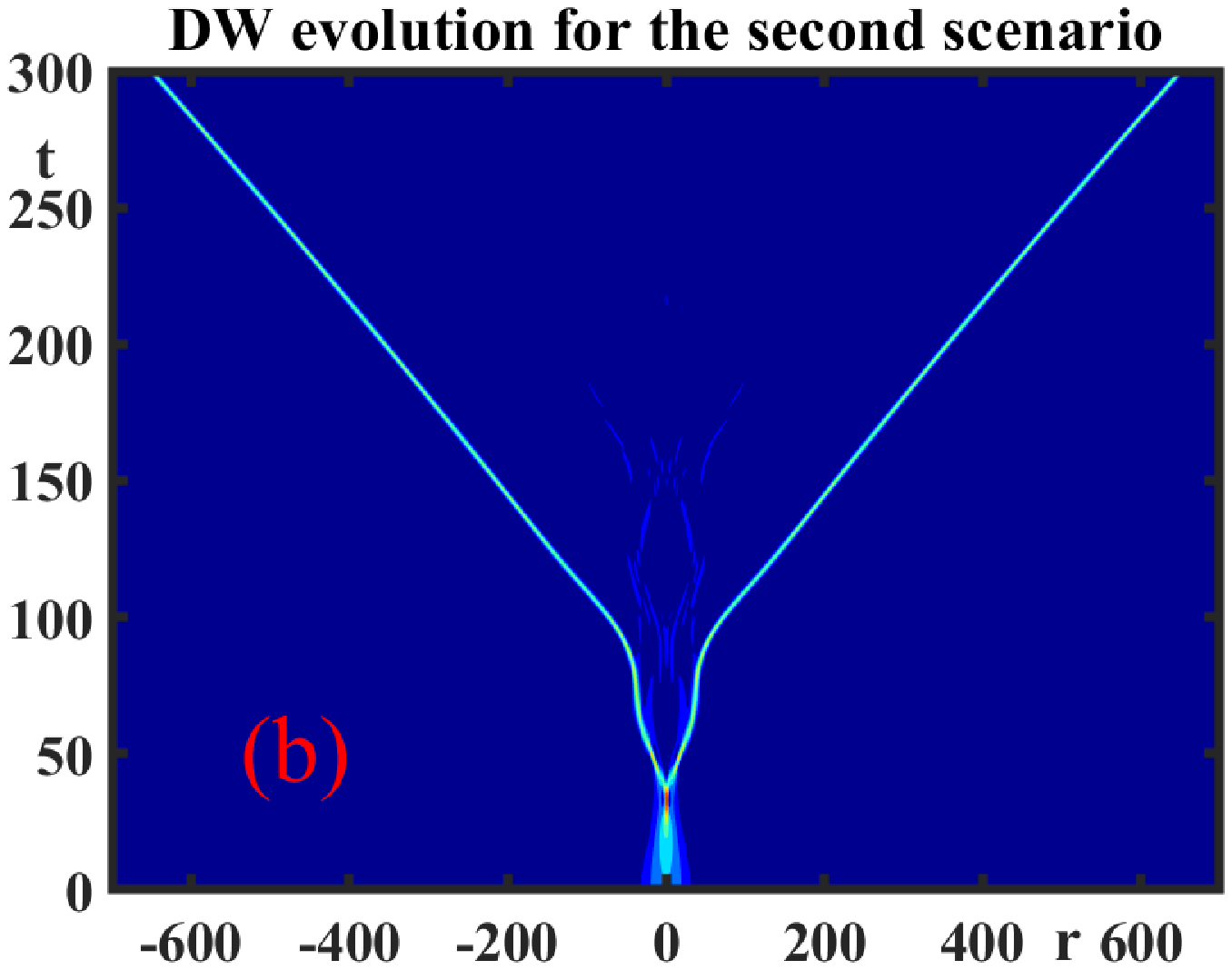}\label{fig2b}}

\subfloat{\includegraphics[scale=0.25]{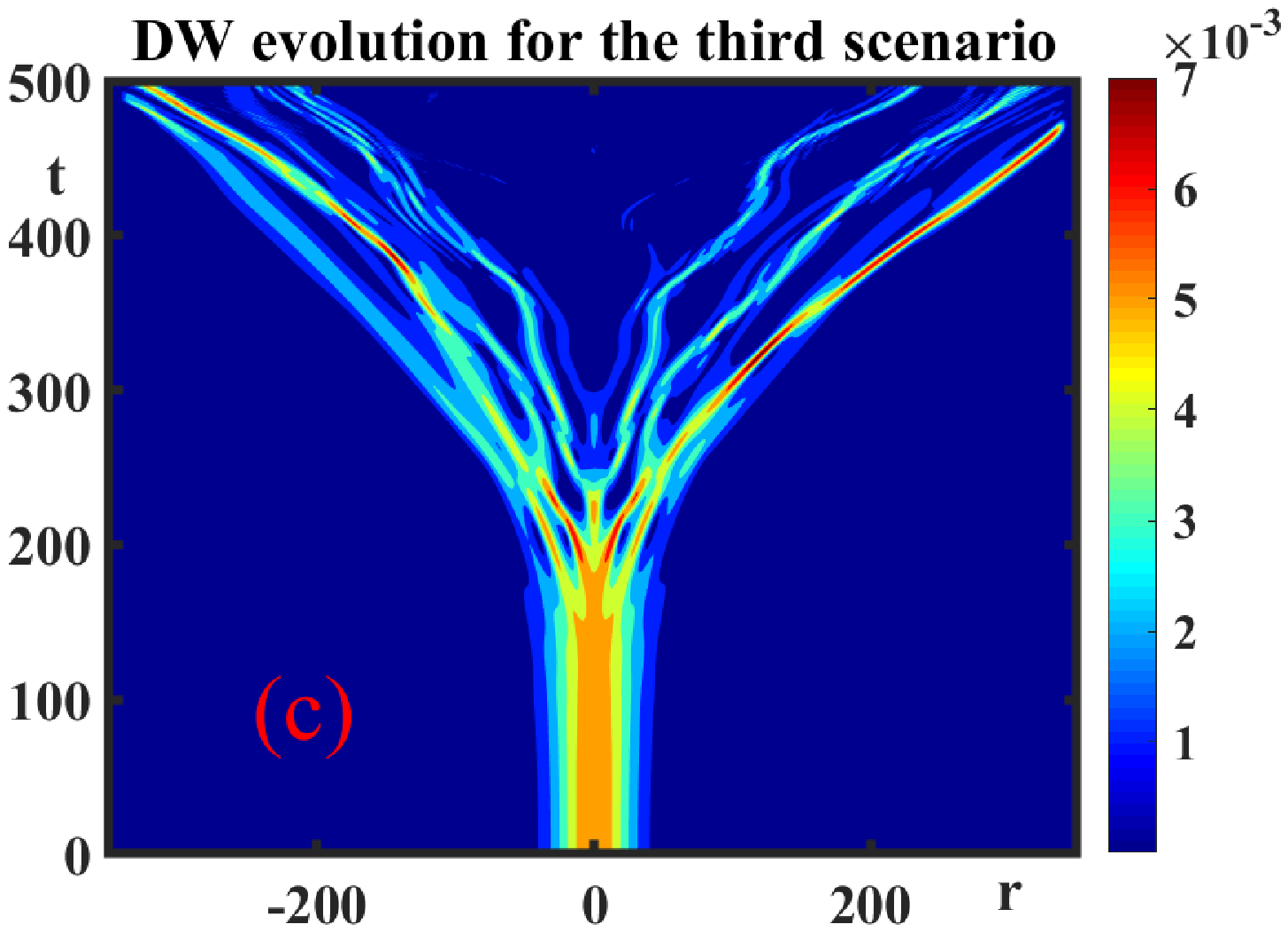}\label{fig2c}}\subfloat{\includegraphics[scale=0.25]{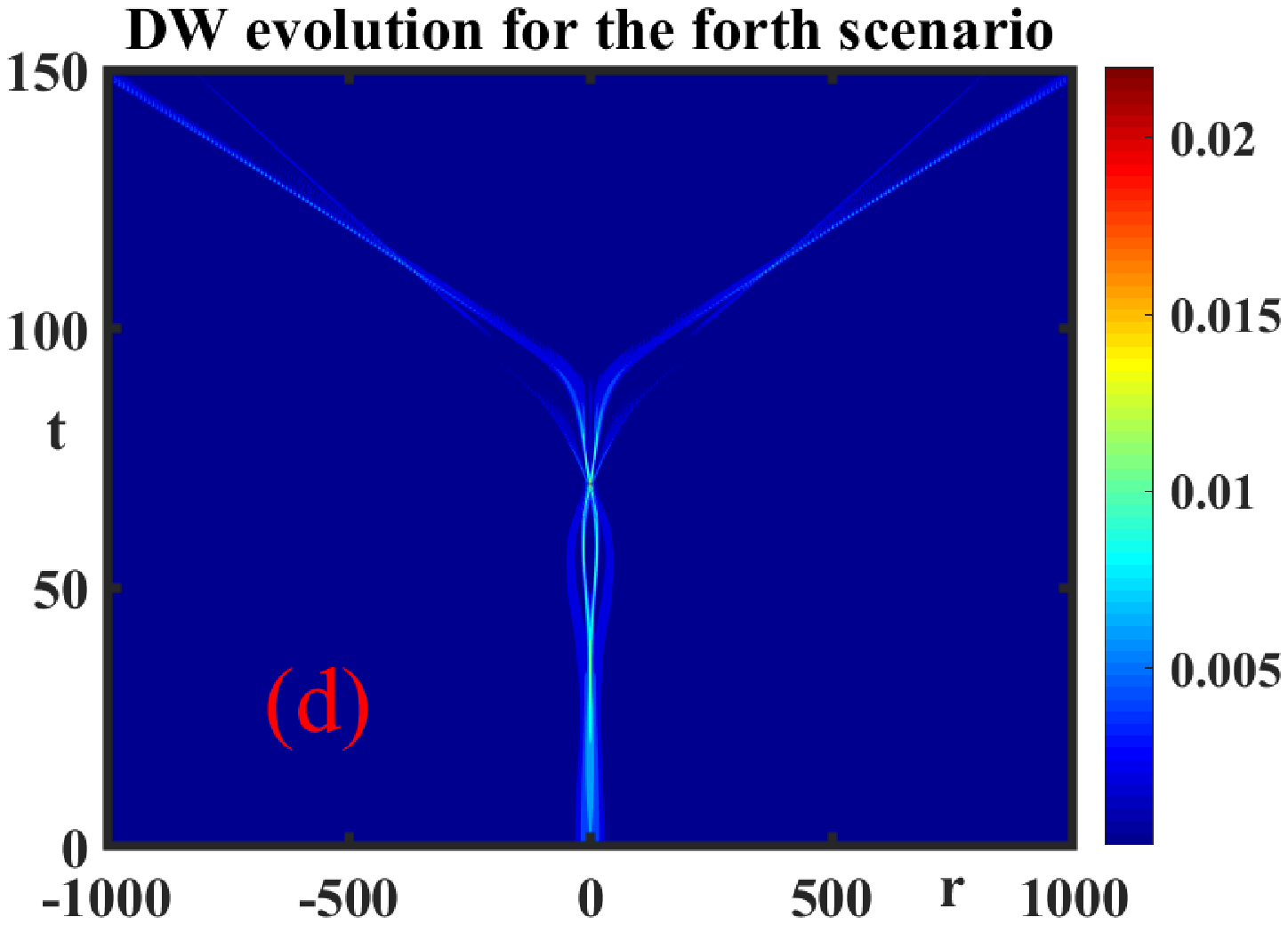}\label{fig2d}}

\caption{The spatial-temporal evolution of the DW for the (a) first scenario,
(b) second scenario, (c) third scenario and (d) forth scenario.  The horizontal axis is the radial scale, and the vertical axis is time. \label{fig2} }
\end{figure}

\par\end{center}

\begin{center}
\begin{figure}
\includegraphics[scale=0.4]{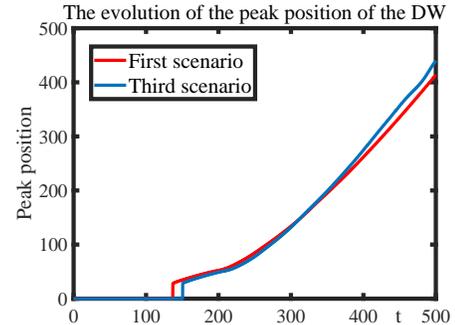}

\caption{The evolution of the radial position of the DW soliton.  The red and blue solid lines represent the peak position of the DW for the first and third scenarios.\label{fig3}}
\end{figure}

\par\end{center}

\begin{center}
\begin{figure}
\subfloat{\includegraphics[scale=0.25]{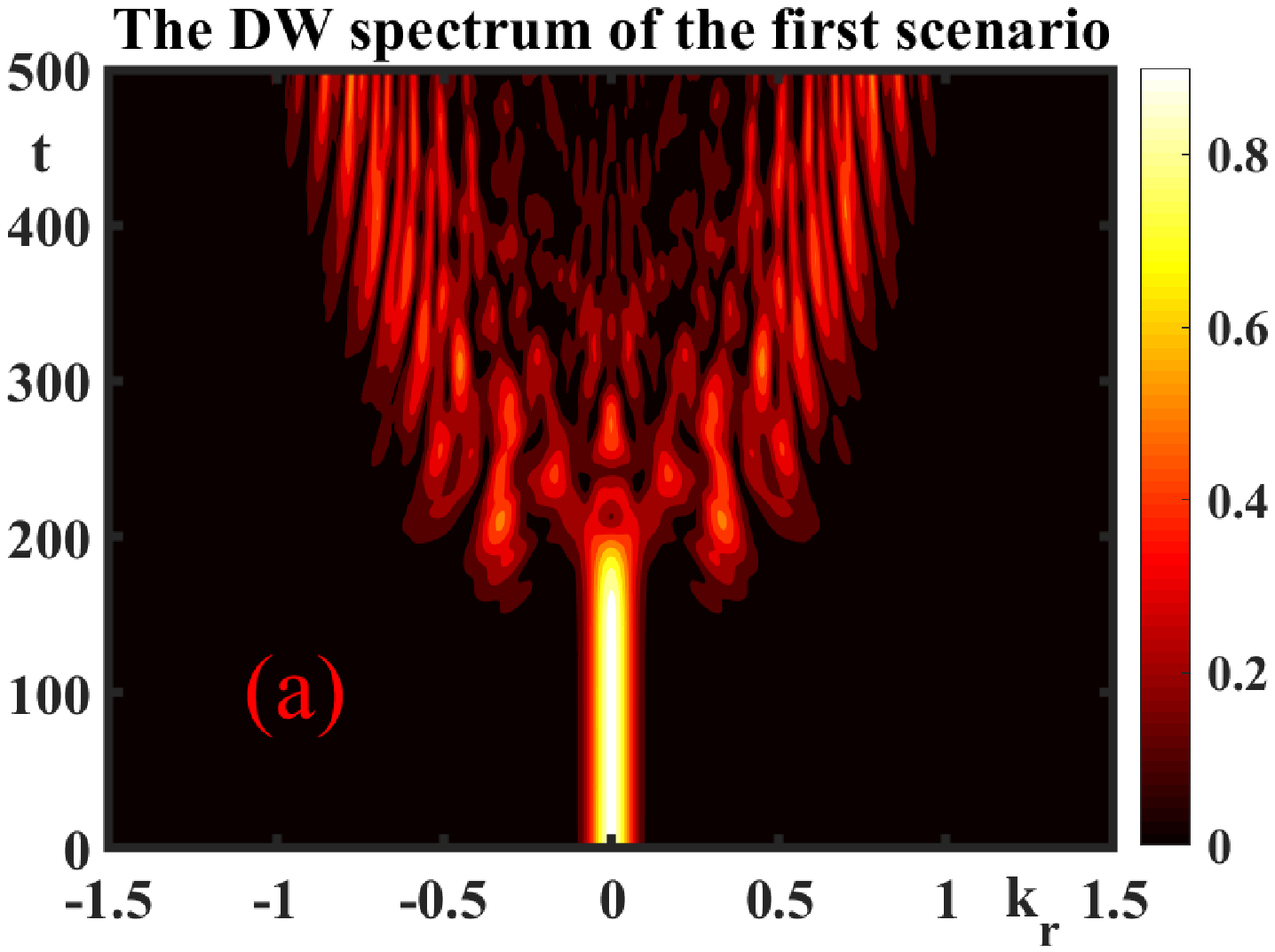}\label{fig4a}}\subfloat{\includegraphics[scale=0.25]{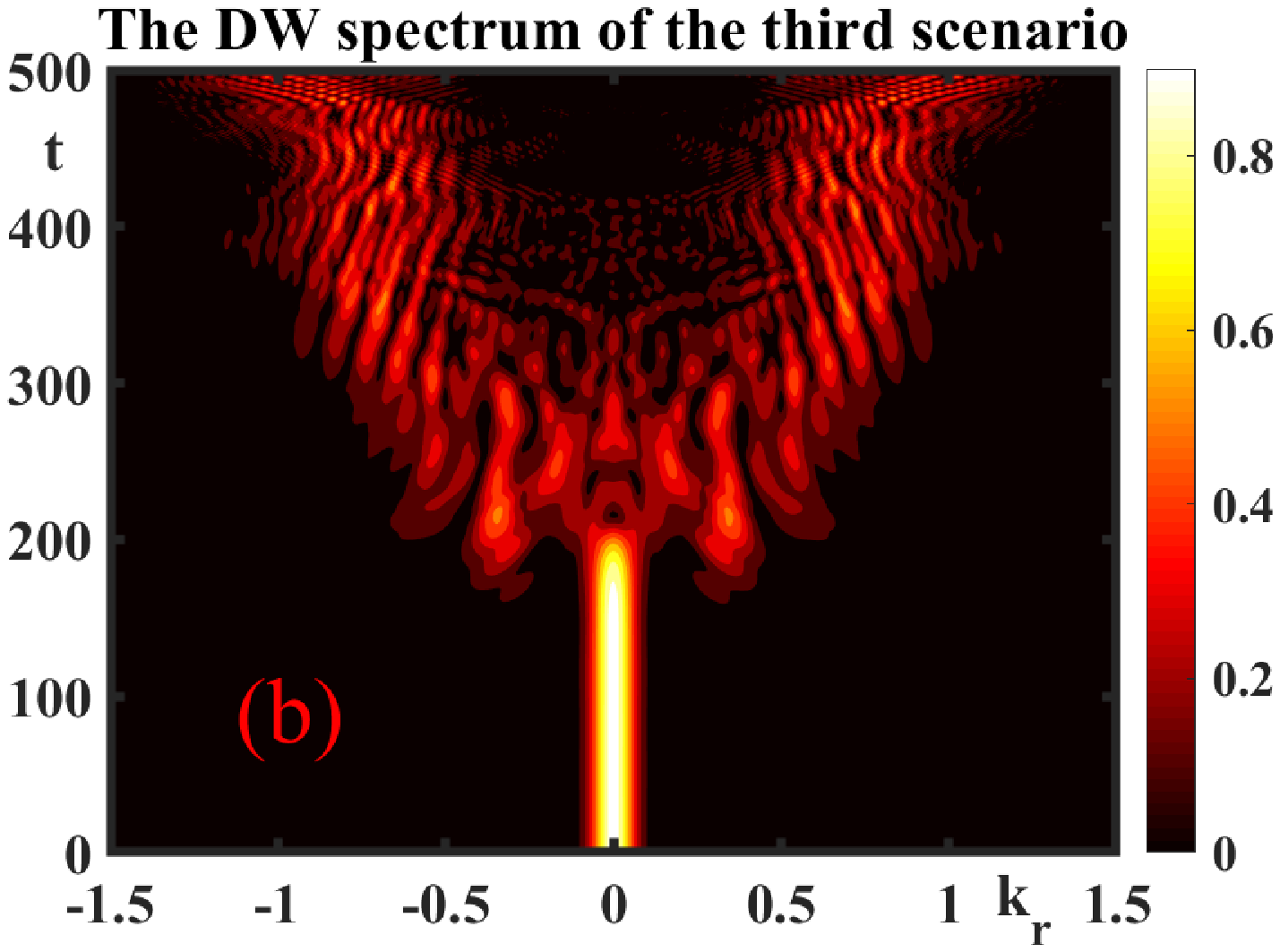}\label{fig4b}}

\subfloat{\includegraphics[scale=0.25]{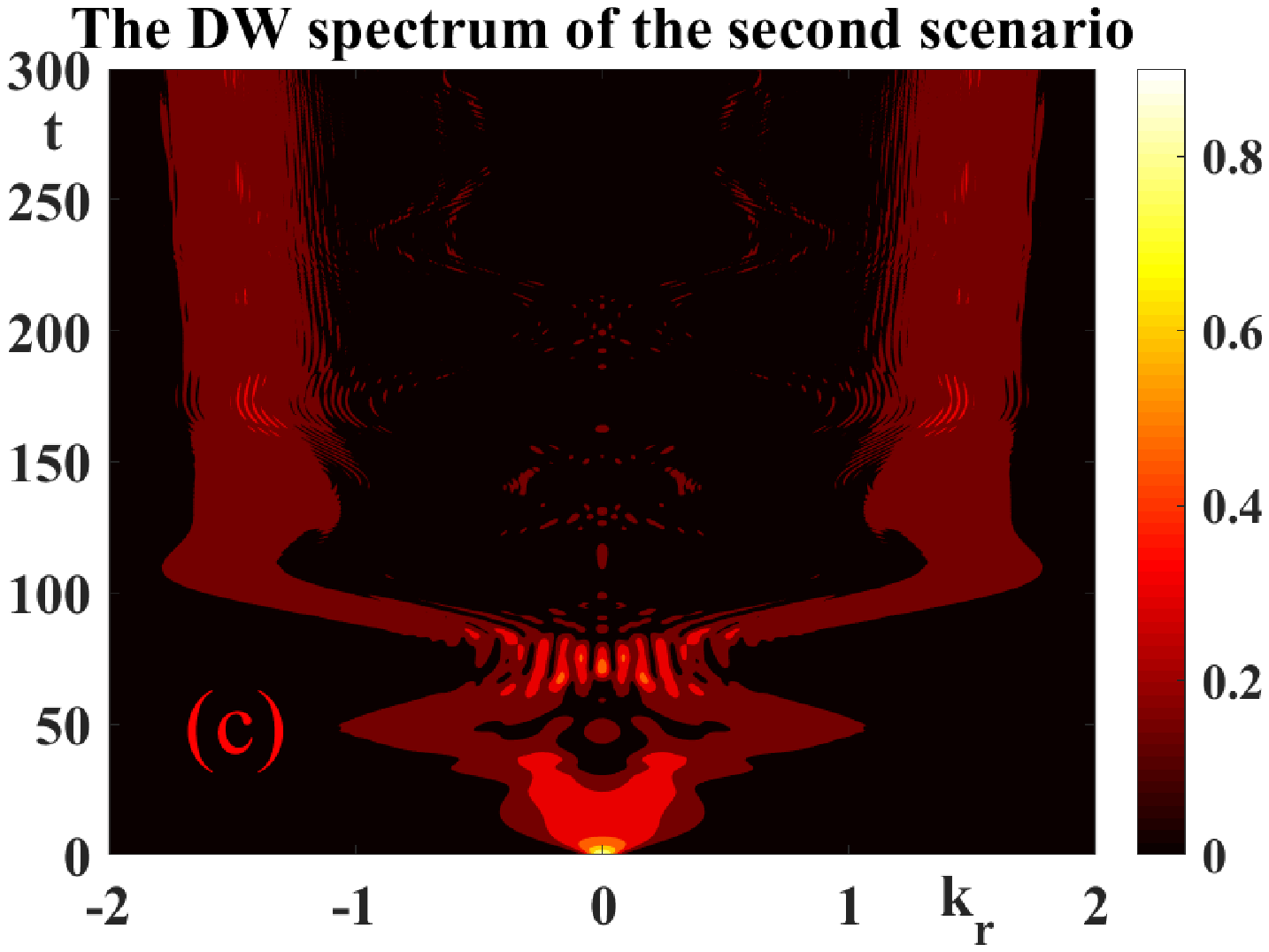}\label{fig4c}}

\caption{The evolution of the $k_{r}$ spectrum of the DW for the (a) first,
(b)third and (c) second scenarios. \label{fig4} }
\end{figure}

\par\end{center}

\begin{center}
\begin{figure}
\subfloat{\includegraphics[scale=0.25]{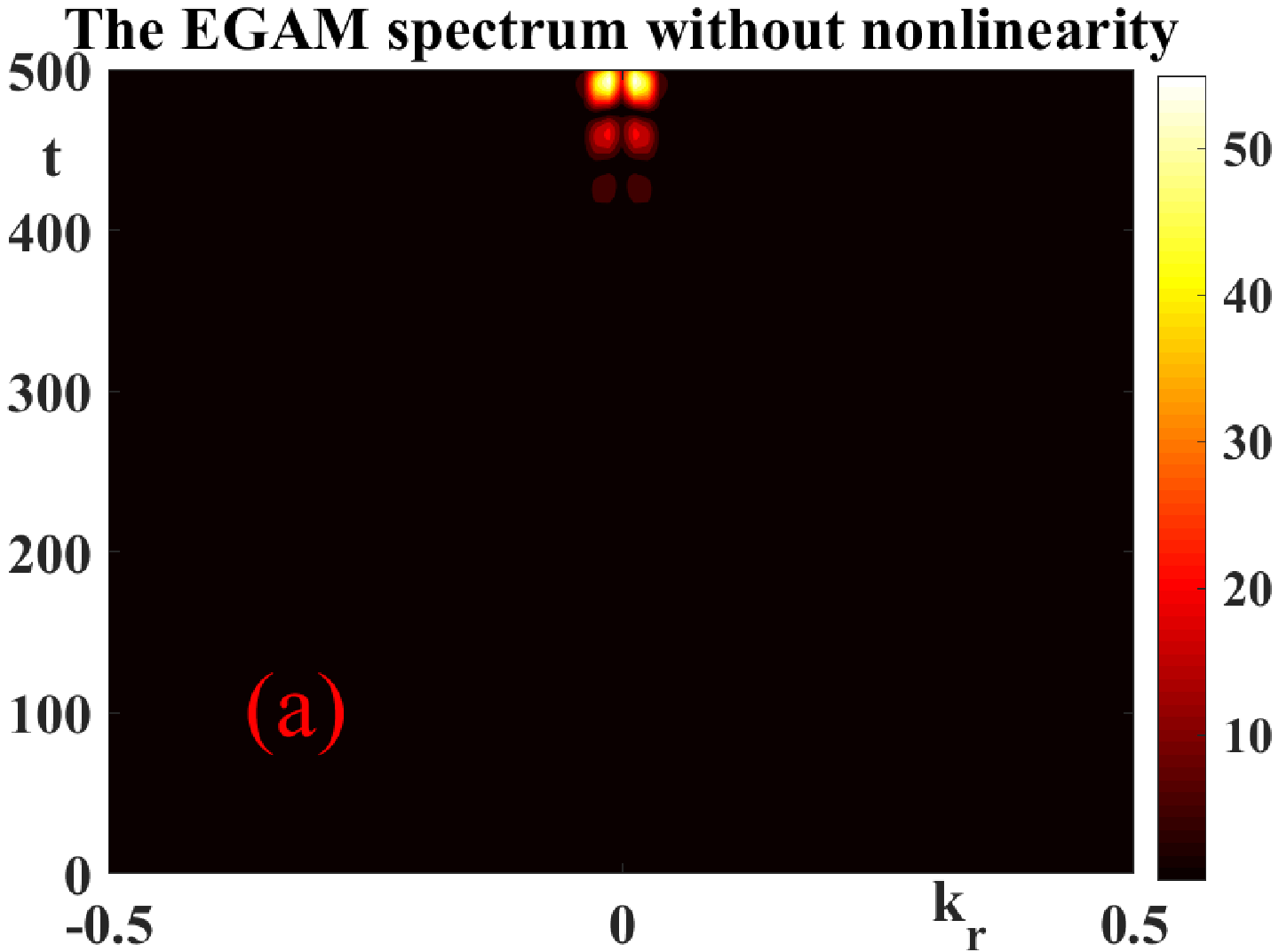}\label{fig5a}}\subfloat{\includegraphics[scale=0.25]{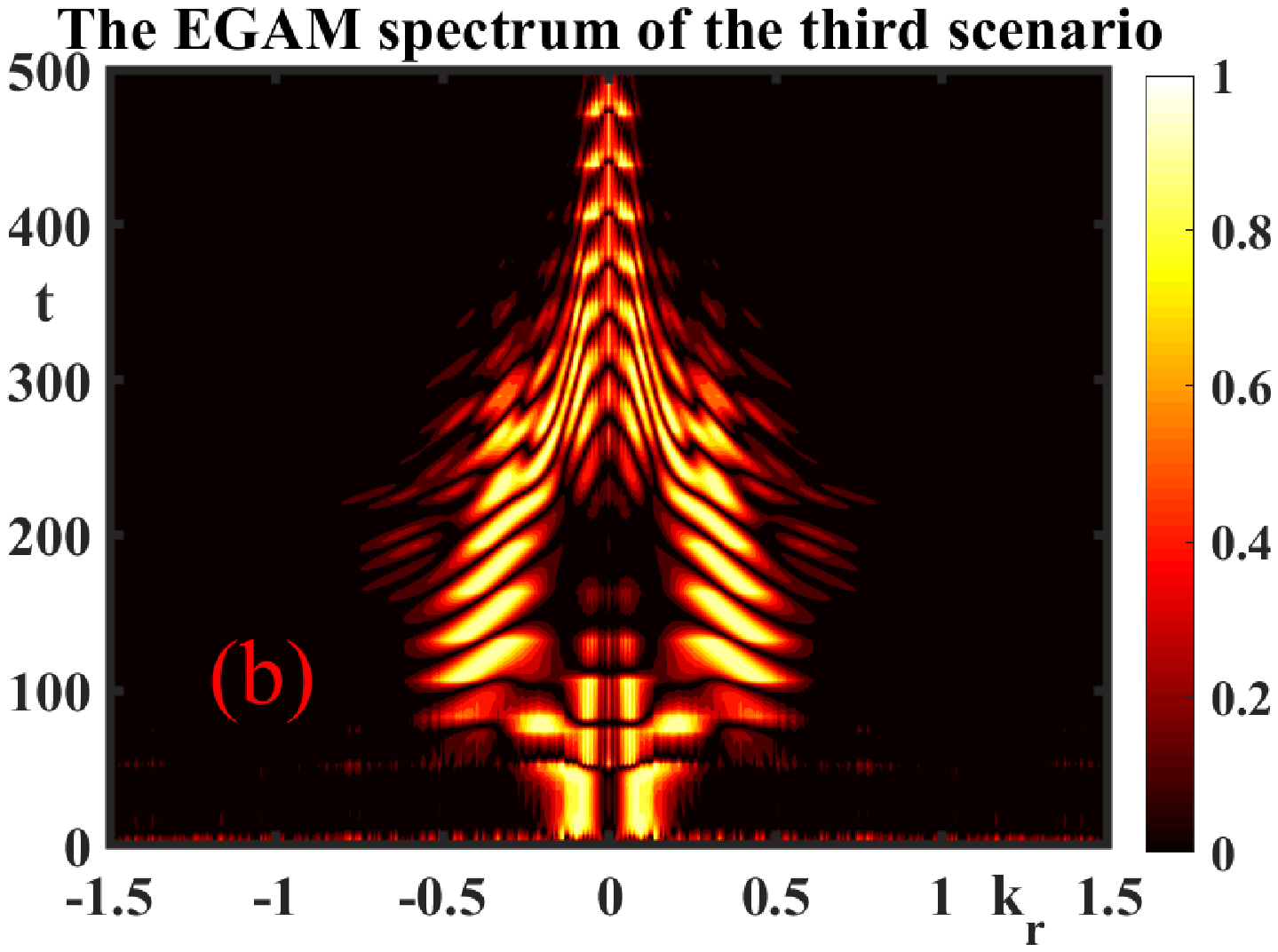}\label{fig5b}}

\subfloat{\includegraphics[scale=0.25]{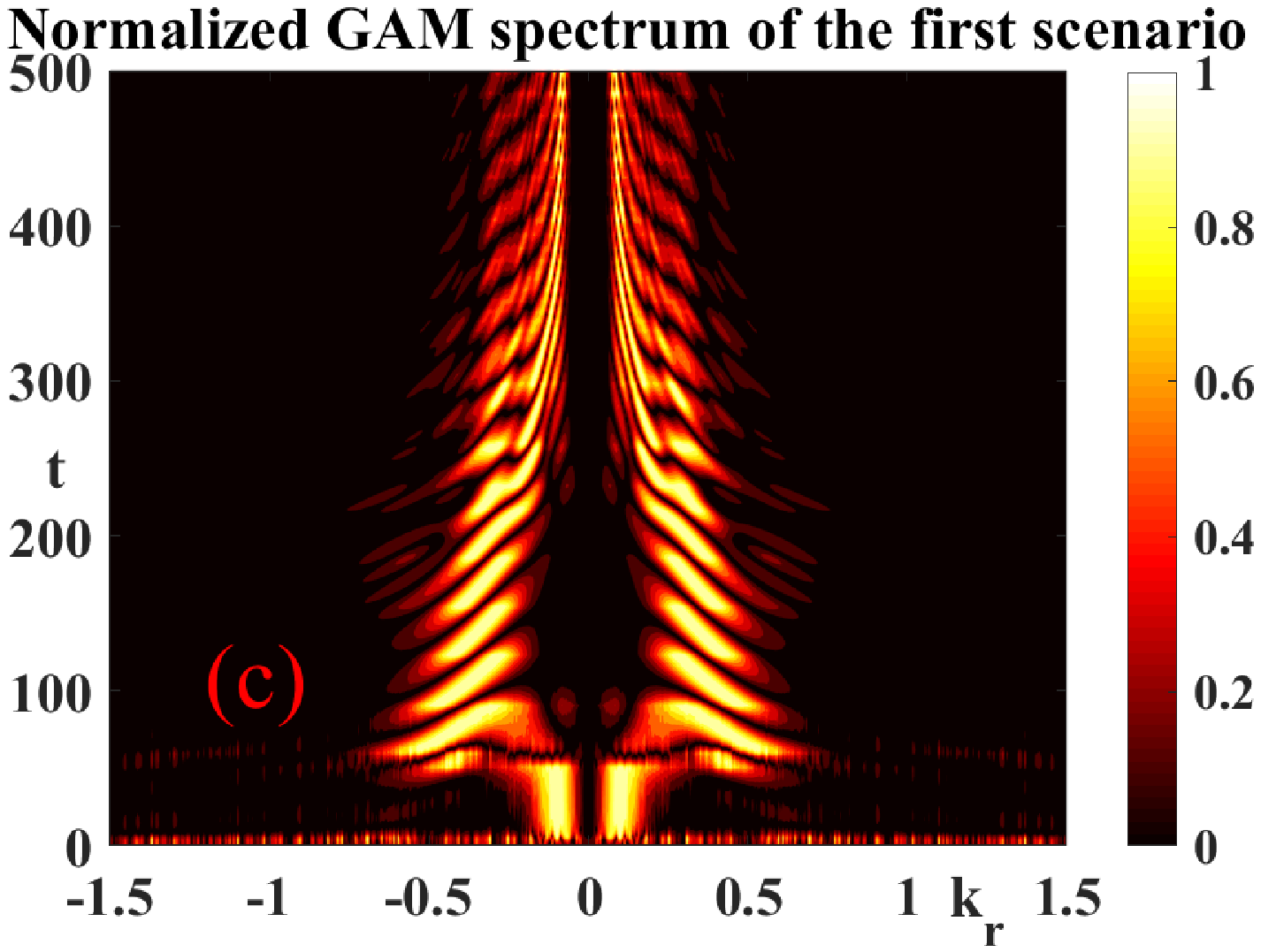}\label{fig5c}}

\caption{The evolution of the $k_{r}$ spectrum of the EGAM for (a) nonlinearity
turned off, (b) nonlinearity turned on and (c) nonlinearity turned
on without the linear EPs drive. The spectrum shown in (b) and (c) are normalized to the maximum at each time step. \label{fig5} }
\end{figure}

\par\end{center}

\section{Conservation laws and energy flow\label{sec:Energy-flow}}

The conservation laws are significant for describing a nonlinear system.
The conservation law can be written as $\rho_{t}+J_{x}=S_{0}$, where
$\rho$ is the ``density", $J$ is the corresponding flux, and $S_{0}$ is the source.
Assuming that as $x\rightarrow\infty$, $J\rightarrow0$, then
general form of a conserved quantity can be written as

\begin{eqnarray}
\dfrac{d}{dt}\int_{-\infty}^{+\infty}\rho dx & = & \int_{-\infty}^{+\infty}S_{0}dx.
\end{eqnarray}

This expression demonstrates that $\rho$ is conserved in the absence
of the source $S_0$. In this section, the conservation laws, especially
the energy conservation law, of the nonlinear DW-EGAM system will
be derived and analyzed to figure out the energy transfer process
of the nonlinear system. The first conservation law can be directly
derived from the nonlinear evolution equation of DW and it has already been derived in the Ref. \cite{NChenPPCF2022}. By adding
$u^{*}\times(\ref{eq:DW Eq})$ to its complex conjugate, the conservation
law can be written as
\begin{eqnarray}
\left|u\right|_{t}^{2}+iC_{d}\left(u^{*}u_{r}-uu_{r}^{*}\right)_{r} & = & 0.\label{eq:1st conserved}
\end{eqnarray}

\begin{center}
\begin{figure}
\includegraphics[scale=0.4]{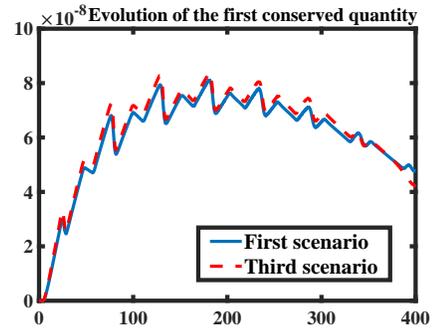}

\caption{The conservation of the first conserved quantity $W$ is shown here, with the  blue solid and red dashed lines representing the temporal evolution of $W$ for the first and third scenarios. \label{fig6}}

\end{figure}

\end{center}

Then, the conserved quantity can be derived by integrating the equation
over the whole domain, i.e., $dW/dt =0$,
where   $W\equiv\left\langle \left|u\right|^{2}\right\rangle$ with $\left\langle \cdots\right\rangle \equiv\int_{-\infty}^{+\infty}\cdots dr$
representing the integration over the whole domain. The conservation
of $W$ is demonstrated
in Fig. \ref{fig6}, which can also be used to benchmark the numerical
scheme. Here, the relative error $(W-W(t=0))/W(t=0)$
is about $10^{-7}$ for the first and third scenarios. The accuracy
of $W$ for the second and forth scenarios is also up
to $10^{-4}$, though it is much larger than that in the first and
third scenario. Based on the first conservation law, equation (\ref{eq:GAM Eq}) can be rewritten as

\begin{eqnarray}
v_{tt}-2\gamma_{G}v_{t}+\omega_{G}^{2}v-C_{G}\omega_{G}^{2}v_{rr} & = & \dfrac{\alpha_{i}\tau\Gamma_{0}}{C_{d}}\left|u\right|_{tt}^{2}.\label{eq:GAM Eq 2}
\end{eqnarray}

The coupled equations (\ref{eq:DW Eq}) and (\ref{eq:GAM Eq 2}) are
similar to the well-known Zakharov equations. Hence, the energy conservation
law can be derived following the standard procedure of the Zakharov
system \cite{VZarharovJETP1972}. By defining $\phi_{t}\equiv v$, the coupled equations with
the form of the conservation laws can be derived by adding $u_{t}^{*}\times(\ref{eq:DW Eq})$
to its complex conjugate and multiplying equation (\ref{eq:GAM Eq 2})
with $\phi$, which can be written as

\begin{eqnarray}
 & \left[\left|u\right|^{2}+\dfrac{C_{d}}{2}\left(uu_{rr}^{*}+u^{*}u_{rr}\right)\right]_{t}+\dfrac{C_{d}}{2}\left(u_{r}u_{t}^{*}+u_{r}^{*}u_{t}\right.\nonumber \\
 & \left.-uu_{tr}^{*}-u^{*}u_{tr}\right)_{r}=-\Gamma_{0}v\left|u\right|_{t}^{2},\label{eq:energy1}\\
 & \left(\phi v_{t}-\dfrac{v^{2}}{2}+\dfrac{\omega_{G}^{2}\phi^{2}}{2}+\dfrac{C_{G}\omega_{G}^{2}\phi_{r}^{2}}{2}-2\gamma_{G}\phi v\right)_{t}\nonumber \\
 & +\left(-C_{G}\omega_{G}^{2}\phi\phi_{rt}\right)_{r}=-\dfrac{\Gamma_{0}\alpha_{i}\tau}{C_{d}}v\left|u\right|_{t}^{2}.\label{eq:energy2}
\end{eqnarray}

Combining the above equations and integrating over the whole domain,
the energy conservation quantity can be obtained

\begin{eqnarray}
\frac{d}{dt}E=\left\langle 2\gamma_Gv^2\right\rangle\label{eq:conservation law},
\end{eqnarray}
with
\begin{eqnarray}
E\equiv \left\langle -{\phi \phi_{tt}+\dfrac{\lvert \phi_t\rvert^{2}}{2}-\dfrac{\omega_{G}^{2}\phi^{2}}{2}-\dfrac{C_{G}\omega_{G}^{2}\lvert\phi_{t}\rvert^{2}}{2}+2\gamma_{G}\phi \phi_t}\right.&&\nonumber\\
\left.{+\dfrac{\alpha_{i}\tau\Gamma_{0}}{C_{d}}\phi\left|u\right|_{t}^{2}}+\alpha_{i}\tau\left({\dfrac{\left|u\right|^{2}}{C_{d}}+\dfrac{\left(u^{*}u_{rr}+uu_{rr}^{*}\right)}{2}}\right)\right\rangle.&&\nonumber
\end{eqnarray}

\begin{center}
\begin{figure}
\subfloat{\includegraphics[scale=0.25]{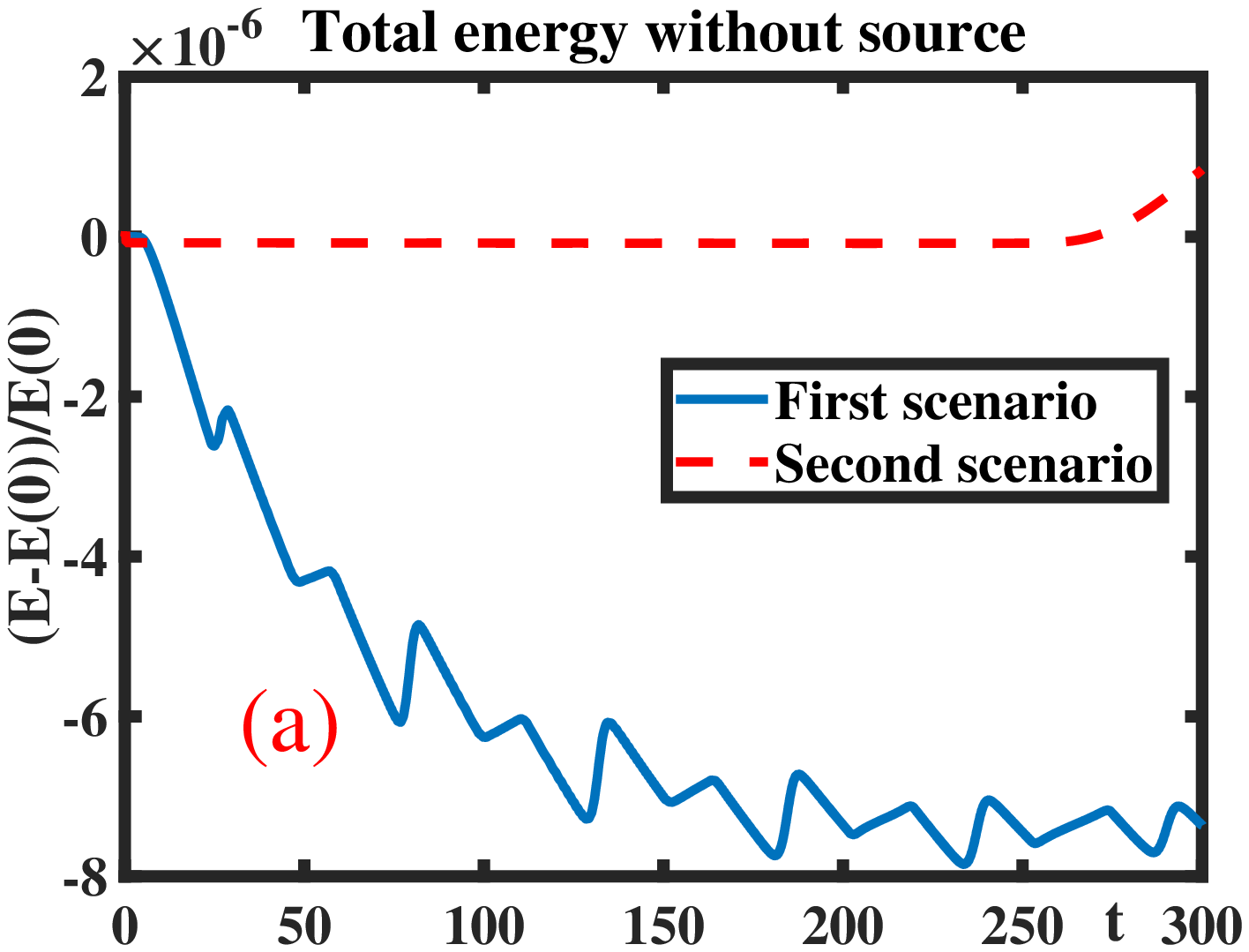}\label{fig7a}}\subfloat{\includegraphics[scale=0.25]{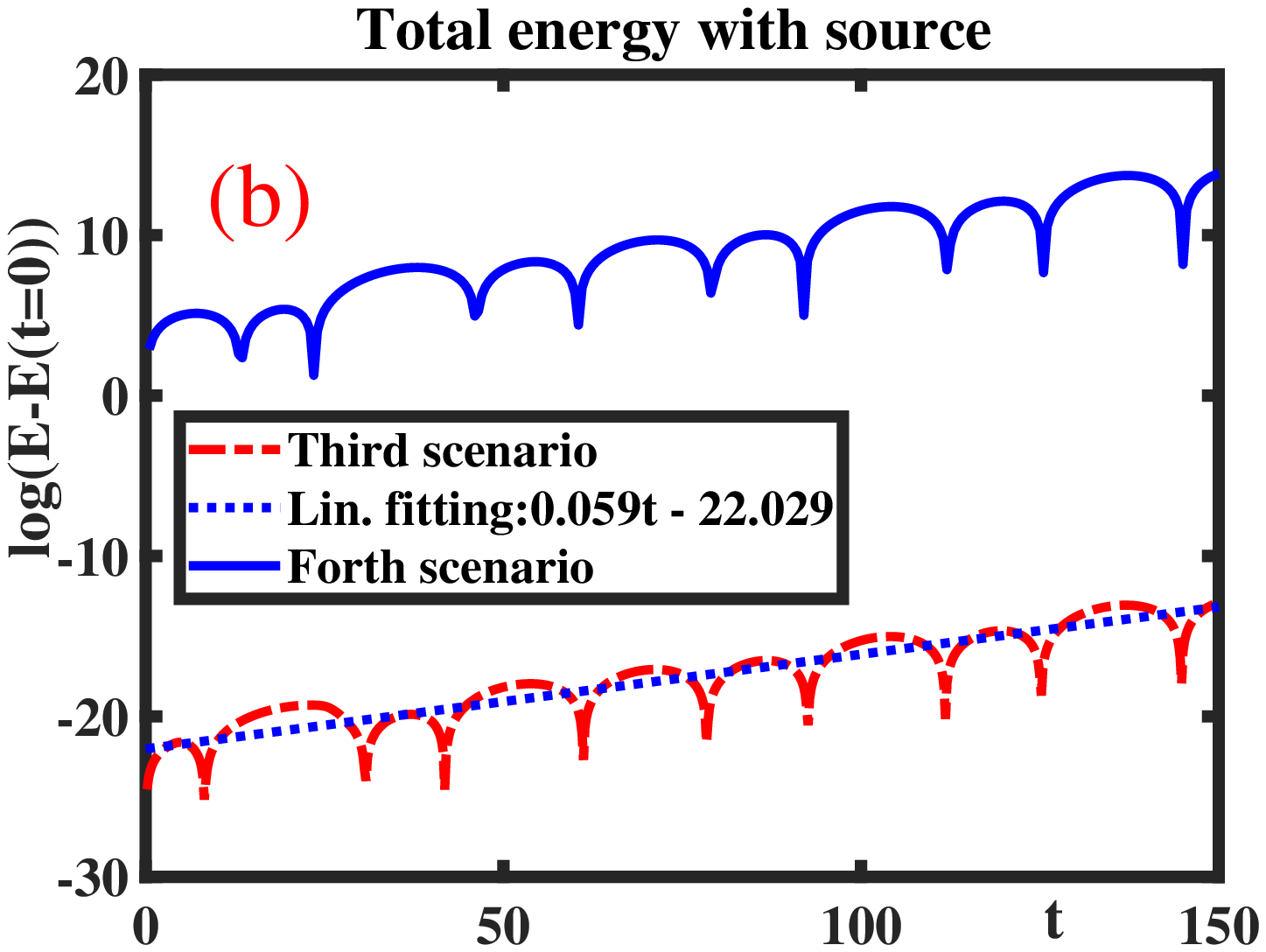}\label{fig7b}}

\caption{The evolution of the total energy for the (a) third and (b) forth
scenarios is shown here, with the blue solid and red dashed lines in (a) representing the evolution of $E$ for the first and second scenarios. The deep blue solid and red dashed lines in (b) demonstrate the evolution of the logarithm of $E$ for the forth and third scenarios, while, the blue dashed line is the linear fitting. \label{fig7} }
\end{figure}
\end{center}

The terms on the LHS of the E represent the time derivative of the total energy
$E$ of the nonlinear system, indicating that the increase of
$E$ originates from the linear energy source on the RHS. Thus, in
the absence of the linear EPs drive, i.e., in the first and second
scenarios, the total energy should be conserved. This is demonstrated in
the Fig. \ref{fig7}, where  the evolution of the relative error
of the total energy $((E-E(t=0))/E(t=0))$ are  presented. It is found that the total energy is accurately conserved
up to $10^{-6}$. Additionally, for the third and forth scenarios with finite linear EPs
drive, the total energy increases exponentially because of the source
term on the RHS of the equation (\ref{eq:conservation law}). Note that,
the RHS of equation (\ref{eq:conservation law}) is proportionate to $\exp(2\gamma_{G0}t)$ through $\left\langle v^2\right\rangle$,
which can be validated if the slope of $\ln E-t$ diagram is $2\gamma_{G}$.
The slope of the total energy logarithm are 0.059 for both scenarios, as shown by the linear fitting in the Fig. \ref{fig7b}, and we can conclude that the total energy $W$ is still conserved
accounting for the energy source. Above all, the total energy is accurately
conserved for four scenarios investigated here, and this fact establishes
a solid foundation for the investigation of energy transfer process.

The ZFs is expected to reduce the turbulence level by the  flow shearing \cite{ZLinScience1998,TSHahmPoP1999}. Thus, the ``deliberate'' excitation of the EGAM by the externally injected EPs is proposed as an active control of  DW turbulence. However, the ``unexpected'' excitation of DW turbulence from the Dimits shift region by the EGAM was observed in the gyrokinetic simulation, indicating possible energy transfer from the EGAM to the DW \cite{DZarzosoPRL2013}. This unusual result can be understood
as an analogy with the le chatelier principle in the chemistry, where
the extra introduction of the product, EGAM in this case, would trigger
the reverse reaction, i.e., the enhancement of the DW turbulence.
Thus, the energy transfer in the
nonlinear system is quite an important topic to clarify the controversy. This is accomplished by investigating the evolution of each wave components. The total energy can be decomposed into three components, i.e., $E=E_G+E_D+E_I$, with $E_G$ and $E_D$ coming from the linear terms of EGAM and DW, respectively, and $E_I$ stemming from the nonlinear terms. More specifically, each energy components can be given explicitly as follows

\begin{eqnarray}
E_G\equiv \left\langle -\phi \phi_{tt}+\dfrac{ \phi_t^{2}}{2}-\dfrac{\omega_{G}^{2}\phi^{2}}{2}-\dfrac{C_{G}\omega_{G}^{2}\phi_{x}^{2}}{2}+2\gamma_{G}\phi \phi_t\right\rangle,& &\nonumber
\end{eqnarray}

\begin{eqnarray}
E_D\equiv \left\langle \alpha_{i}\tau\left(\dfrac{\left|u\right|^{2}}{C_{d}}+\dfrac{\left(u^{*}u_{rr}+uu_{rr}^{*}\right)}{2}\right)\right\rangle,&&\nonumber
\end{eqnarray}

\begin{eqnarray}
E_I\equiv\left\langle\dfrac{\alpha_{i}\tau\Gamma_{0}}{C_{d}}\phi\left|u\right|_{t}^{2}\right\rangle.&&\nonumber
\end{eqnarray}

The energy transfer is demonstrated by the evolution of each energy
component, i.e., the decrease/increase of an energy component means
lose/gain energy from other components. The evolution of each energy
component is shown in  Fig. \ref{fig8}. The first scenario is
presented in  Fig. \ref{fig8a}, which corresponds to the
spontaneous excitation of the GAM by the DW. It clearly shows the
transfer of energy from the DW to the GAM, which coincides with the
previous investigations. Fig. \ref{fig8b} shows the evolution of
each energy components in the second scenario. It is found that the
DW pumps energy into the EGAM as that in the first scenario, with the difference  that the interaction energy increases
and decreases periodically in this case. The energy transfer process
for the third and forth scenario are shown in the Fig. \ref{fig8c}
and \ref{fig8d}, both demonstrating the energy transfer from the DW
to the EGAM. The energy of the EGAM $E_G$ is not shown here,  because the exponentially
growing $E_G$ would dominate $E_D$ and $E_I$ due to the total energy conservation. In all four scenarios,  the DW energy always decreases, showing no energy transfer from GAM/EGAM to DW. Thus, we conclude, in the present model for GAM/EGAM interaction with DW, there is no sign of DW ``excitation" by GAM/EGAM, and further investigations are needed to understand the observations of Ref. \citenum{DZarzosoPRL2013}.

\begin{center}
\begin{figure}
\subfloat{\includegraphics[scale=0.25]{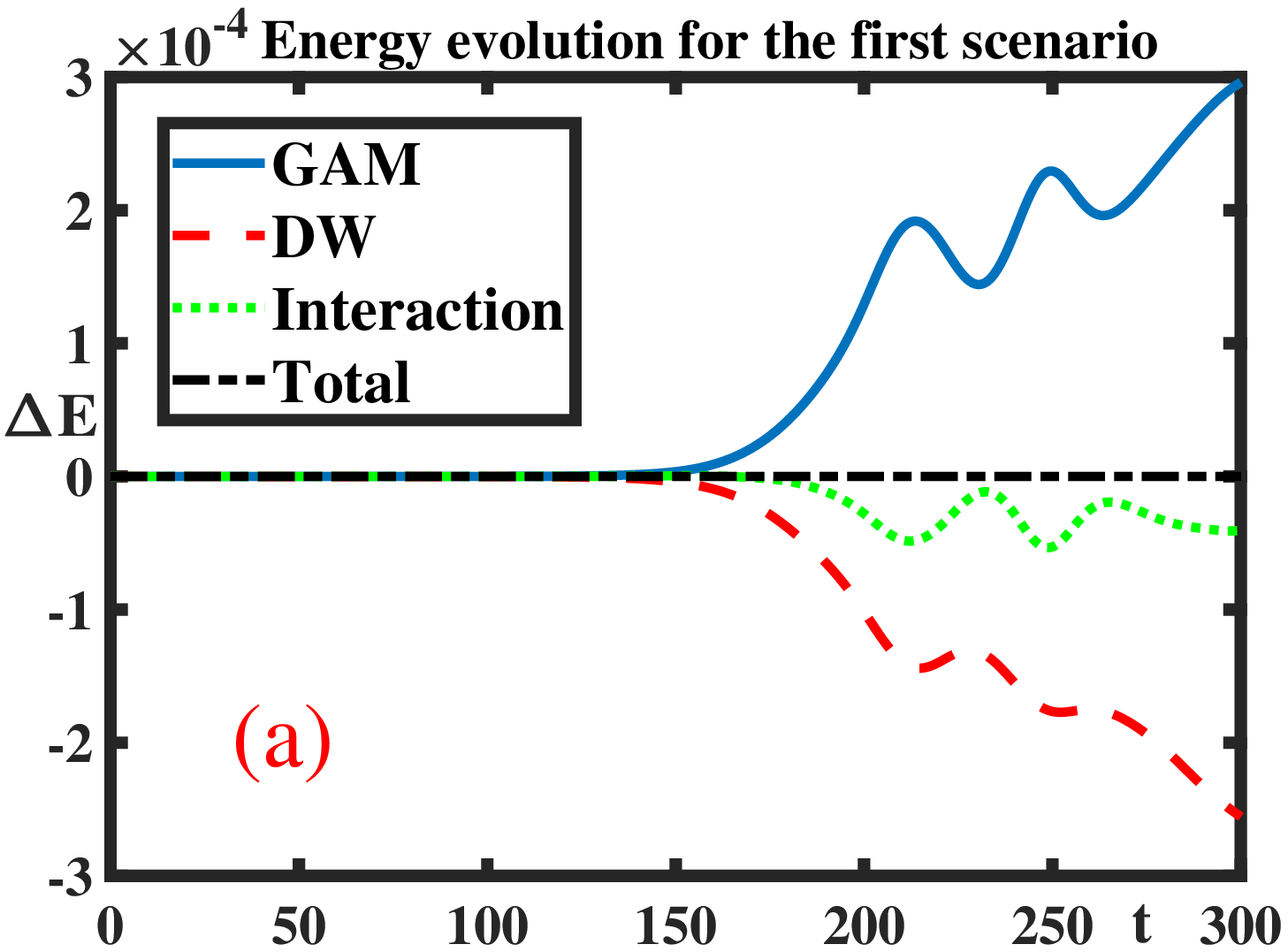}\label{fig8a}}\subfloat{\includegraphics[scale=0.25]{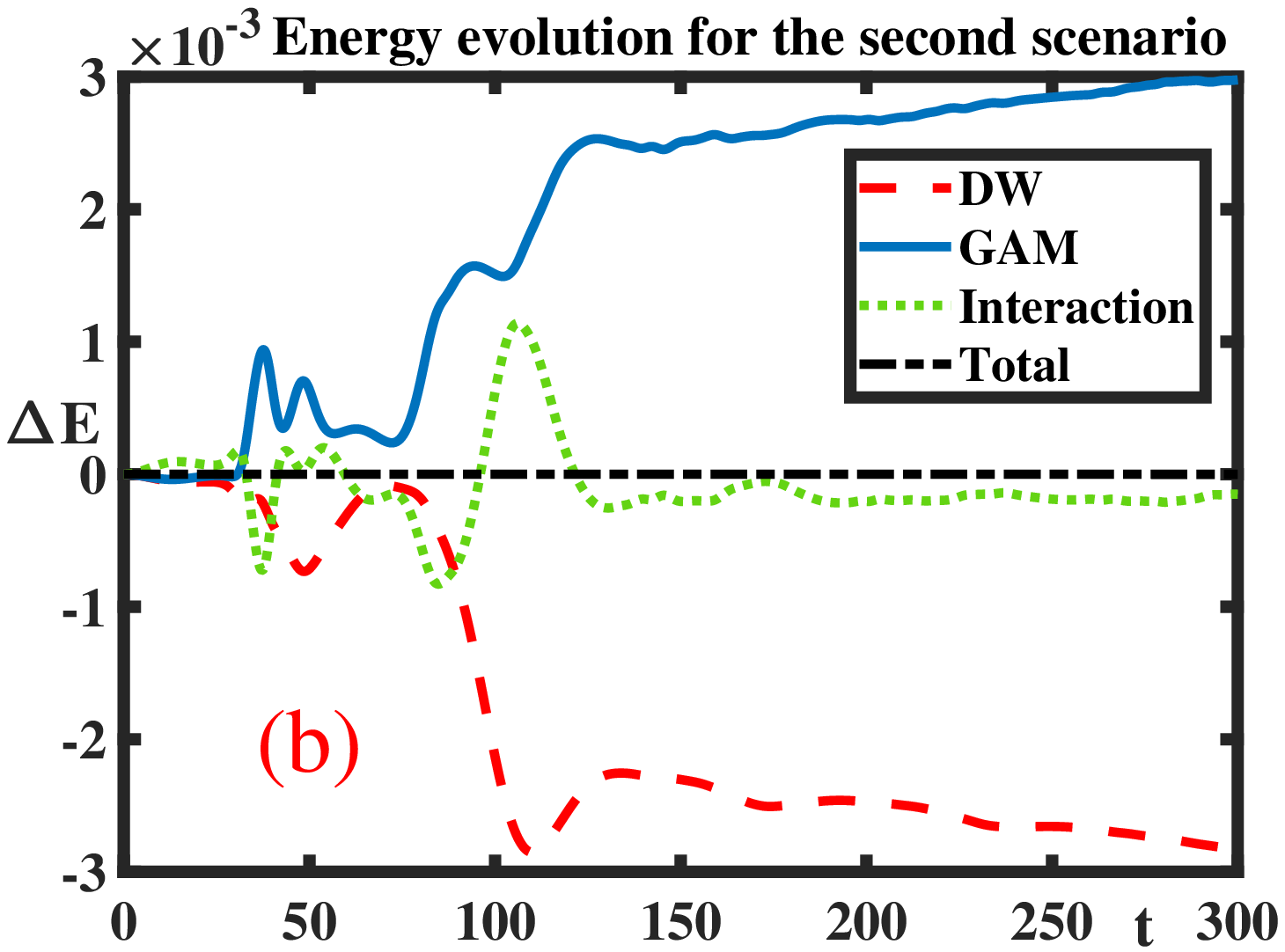}\label{fig8b}}

\subfloat{\includegraphics[scale=0.25]{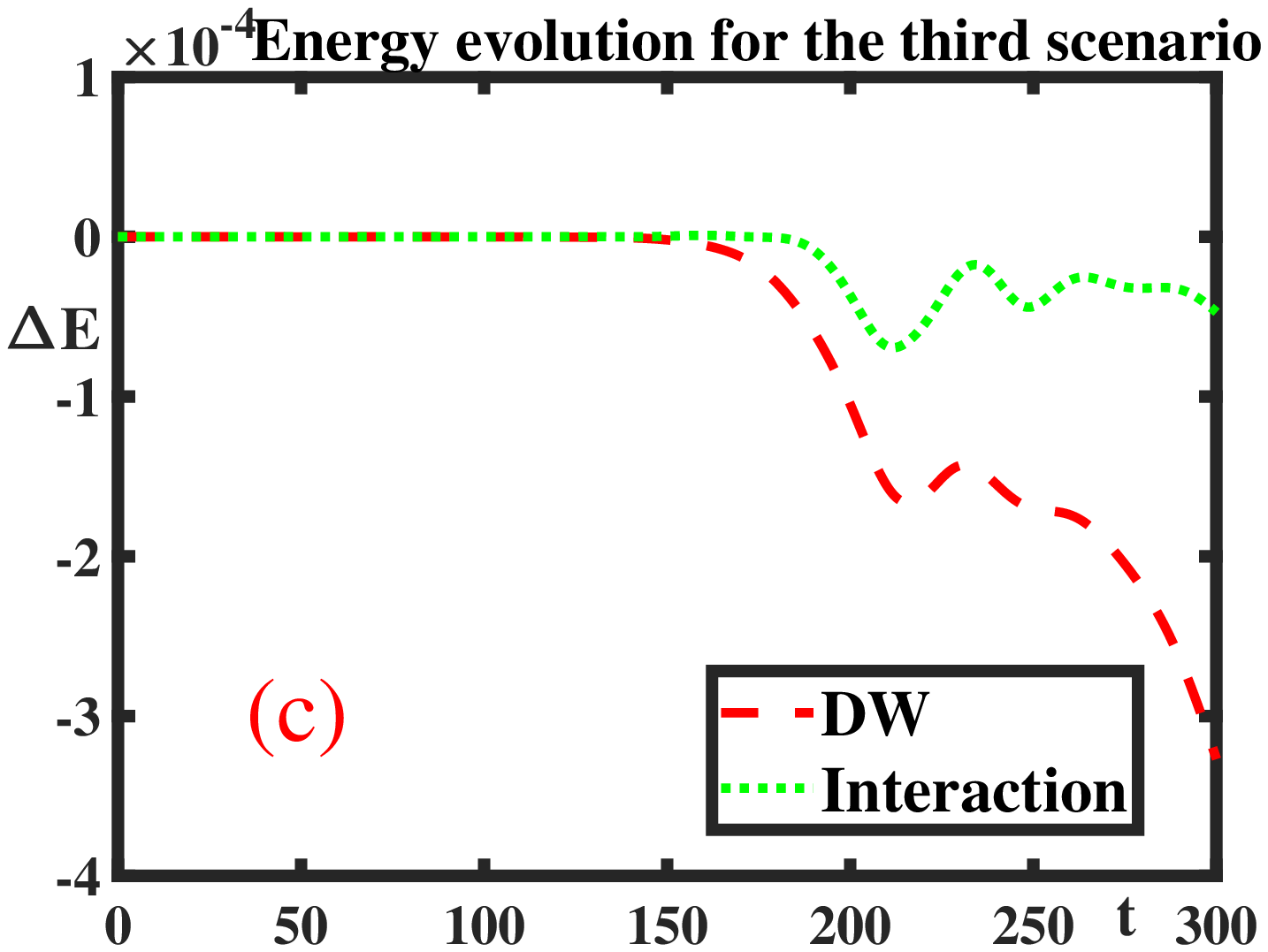}\label{fig8c}}\subfloat{\includegraphics[scale=0.25]{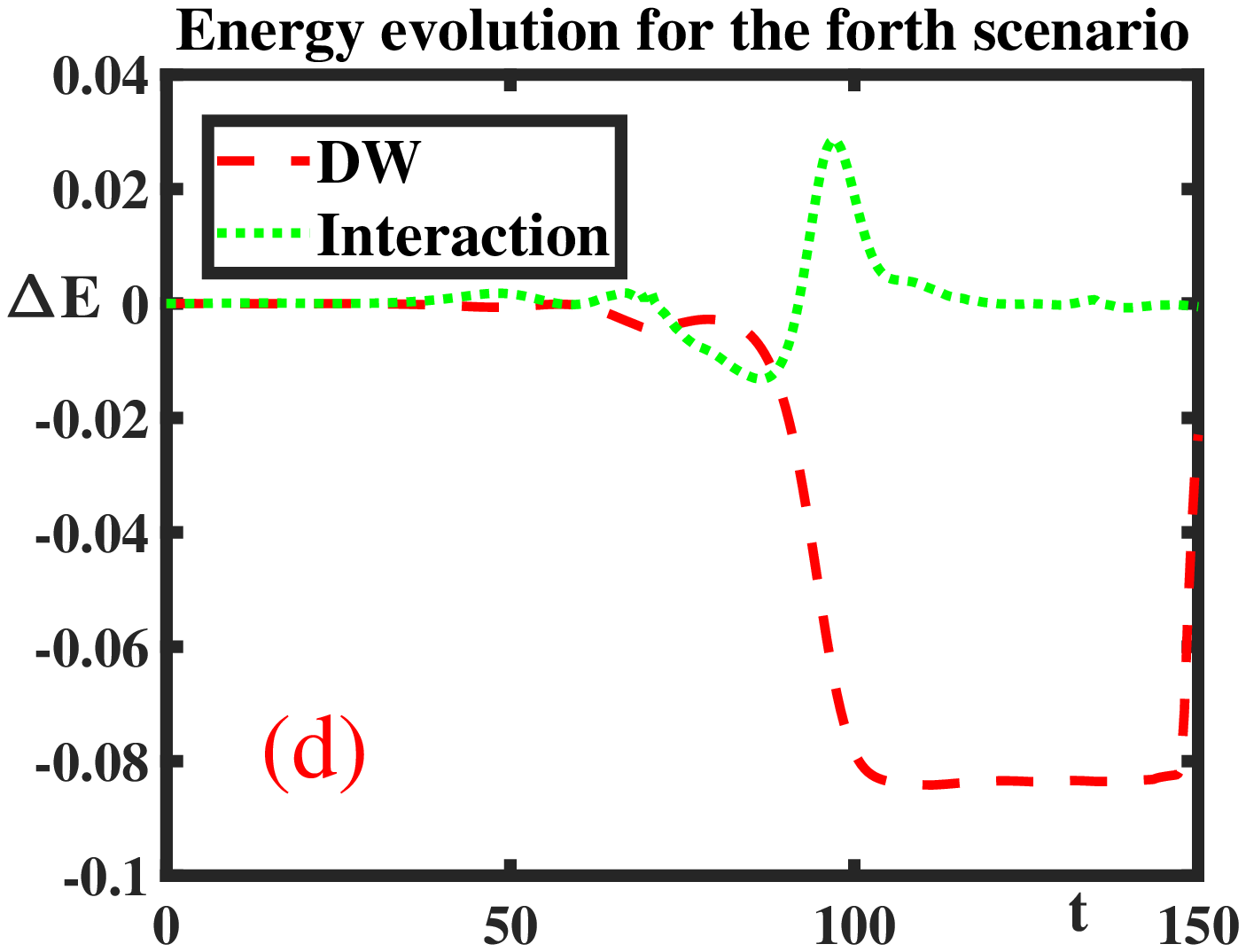}\label{fig8d}}

\caption{The evolution of each energy components for the first to forth scenarios. In the first two figures, the blue solid, red dashed, green dashed and the black dashed lines represent the energy variation of  the EGAM/GAM, DW, Interaction and total, respectively. The deep-blue solid and red dashed lines in the latter figures represent the evolution of the DW and  interaction energy. \label{fig8}}

\end{figure}

\par\end{center}

\section{Conclusion and discussion}\label{conclusion}

In this work, we have investigated the nonlinear interaction between
 ambient DW turbulence and EGAM, by using the first-principle
based two-field equations. Four scenarios with different combinations of EGAM/GAM initial amplitude and linear GAM growth rate are designed to delineate the effects of linear drive and finite GAM amplitude \cite{NChenPoP2021} on DW-EGAM nonlinear interactions \cite{DZarzosoPRL2013}, including DW regulation as well as turbulence spreading.
The energy conservation
law for the nonlinear system is derived and verified numerically.
Based on that, the energy transfer between different energy components
are analyzed in detail.

Firstly, the propagation of DW-EGAM solitons was investigated.
The continuous acceleration of DW-EGAM solitons are observed in
the numerical simulation, in which the peak position of the DW is
presented. The fact may arise from the excitation of higher $k_{r}$
DWs, i.e., the modes with micro-scale mode structures, due to the
modifications of the frequency/wavenumber matching condition by linear EPs drive, resulting in the up-shift of resonant $k_r$. Secondly, two conservation laws, including the energy
conservation law, are derived following that of the nonlinear Zakharov
system. The numerical simulations have proved that the energy is accurately
conserved. In addition, the total energy can be decomposed into the
energy of the DW, EGAM and interaction, thus, the energy transfer
between different modes can be investigated. It is found that
DW transfers energy to EGAM in all four different scenarios, aligning
with common speculation. This result imply that, in this nonlinear DW-EGAM system, GAM/EGAM
stabilizes DW turbulence via the nonlinear energy transfer, as a complement of the previous publications.

Though DW is  damped by EGAM nonlinearly, the enhancement of turbulence spreading from the DW linearly stable to unstable region might offset the positive effect of
EGAM on plasma confinement. The overall outcome needs further investigation.
Moreover, part of the dynamical behaviour of the coupled equations
are surveyed numerically in this work, which is a brief dipiction
of the complex nonlinear system. Thus, more thorough investigations,
especially theoretical treatment, are needed to obtain a better understanding
to the nonlinear system. More importantly, little energy exchanges,
between the meso-scale GAM  and the macro-scale component of EGAM observed in this work,  implies that the indirect nonlinear interactions between
the Alfv\'en eigenmodes (AEs) and the DW, mediated  by the ZFs, might
be relatively weak. Previous works based on the two prey-one predator
model, in which the spatial evolution is neglected, could, thus, be incomplete. As a consequence, the direct cross-scale interactions
of the AEs and DWs \cite{LChenNF2022} could be crucial. Finally, as the linear EPs drive may contribute
to    DW turbulence
spreading,   the ion diamagnetic frequency nonuniformity $\omega_{*}(r)$,  could be  more
significant in the presence of linear EPs drive, to determine the range of turbulence spreading via rendering the convective nature of DW-GAM system into a quasi-absolute instability, as shown in   Ref. \cite{ZQiuPoP2014}. These aspects
will be investigated in a future publication.

\section*{Data availability}

The data that support the findings of this study are available from the corresponding author upon reasonable request.

\section*{Acknowledgement}
This work is supported by the National Science Foundation of China under Grant No.  11875233,  the National Key Research and Development Program of China under Grant No. 2017YFE0301900,   ``Users of Excellence program of Hefei Science Center CAS under Contract No. 2021HSC-UE016", and the Fundamental Research Fund for Chinese Central Universities under Grant No. 226-2022-00216.  The authors acknowledge the fruitful discussion with Prof. Liu Chen (Zhejiang University and University of California at Irvine) and Dr. Fulvio Zonca (Center for Nonlinear Plasma Science, Italy).

\end{document}